\documentclass[12pt,a4paper]{article}
\usepackage[latin1]{inputenc}
\usepackage{a4wide}
\usepackage{amsfonts}
\usepackage{amssymb}
\usepackage{amsmath}
\usepackage{bbm}
\usepackage{caption}
\usepackage{ifpdf}
\ifpdf
\usepackage[pdftex]{graphicx}
\usepackage[pdftex,unicode,implicit]{hyperref}
\hypersetup{%
  pdftitle    = {The general d=5,6 tensor hierarchies},
  pdfkeywords = {Supersymmetry, supergravity, symmetry, gauge},
  pdfauthor   = {Jelle Hartong and Tomás Ortín},
  pdfcreator  = {pdf\LaTeXe\ with package \flqq hyperref\frqq},
  pdfproducer = {pdf\LaTeXe\ with package \flqq hyperref\frqq},
  pdfpagemode = None,  
  pdffitwindow= true,  
  unicode     = true,
  plainpages  = true,
  colorlinks  = true,  
  citecolor   = blue,
  urlcolor    = red,
  linkcolor   = black
}
\newcommand{\hepth}[1]{arXiv:{\tt
\href{http://www.arXiv.org/abs/hep-th/#1}{hep-th/#1}}}

\newcommand{\arxiv}[1]{{\tt
\href{http://www.arXiv.org/abs/#1}{arXiv:#1}}}
\else
  \usepackage[dvips]{graphicx}
  \usepackage[unicode,implicit]{hyperref}
  \newcommand{\hepth}[1]{arXiv:{\tt hep-th/#1}}

  \newcommand{\arxiv}[1]{{\tt arXiv:#1}}
\fi
\makeatletter
\@addtoreset{equation}{section}
\makeatother

\makeindex
\begin{document}

\begin{flushright}
\small
IFT-UAM/CSIC-09-22\\
June $23^{\rm rd}$, 2009\\
\normalsize
\end{flushright}

\begin{center}

\vspace{.7cm}

{\LARGE {\bf Tensor Hierarchies of 5- and 6-Dimensional\\[.5cm] Field Theories}}

\vspace{2.5cm}

\begin{center}

{\bf Jelle Hartong ${}^{\dagger}$ and Tom\'as Ort\'{\i}n} ${}^{\ddagger}$ 

\vspace{.7cm}

${}^{\dagger}$ {\em Institute for Theoretical Physics, \\
Sidlerstrasse 5, 3012 Bern,
Switzerland\vskip 5pt}

{e-mail: {\tt hartong@itp.unibe.ch}}
\vskip 15pt

${}^{\ddagger}$ {\em Instituto de F\'{\i}sica Te\'orica UAM/CSIC
Facultad de Ciencias C-XVI, \\
C.U. Cantoblanco, E-28049-Madrid, Spain\vskip 5pt}

{e-mail: {\tt Tomas.Ortin@uam.es}}

\end{center}

\vspace{2.5cm}

{\bf Abstract}

\begin{quotation}

  {\small 
    We construct the tensor hierarchies of generic, bosonic, 5- and
    6-dimensional field theories. The construction of the tensor hierarchy
    starts with the introduction of two tensors: the embedding tensor
    $\vartheta$ which tells us which vector is used for gauging and another
    tensor $Z$ which tells us which vector is eaten by a 2-form. In dimensions
    $d\ge 5$ these two (deformation) tensors are in principle
    unrelated. Besides $\vartheta$ and $Z$ there can be further deformation
    tensors describing other couplings unrelated to (but compatible with)
    gauge symmetry. For each deformation tensor there appears a $(d-1)$-form
    potential and for each constraint satisfied by the deformation tensors
    there appears a $d$-form potential in the tensor hierarchy. For each symmetry of the
    undeformed theory there is an associated
    $(d-2)$-form appearing in the tensor hierarchy. Our methods easily
    generalize to arbitrary dimensions and we present a general construction
    for the $d$-, $(d-1)$- and $(d-2)$-form potentials for a tensor hierarchy
    in $d$ dimensions.
}

\end{quotation}

\end{center}

\newpage
\pagestyle{plain}

\tableofcontents

\newpage

\section{Introduction}

The structure of the tensor hierarchy\footnote{Tensor hierarchies have been
  introduced in Refs.~\cite{deWit:2005hv,deWit:2005ub,deWit:2008ta}. They
  arise naturally in the embedding tensor formalism
  \cite{Cordaro:1998tx,deWit:2002vt,deWit:2003hq,deWit:2005hv,deWit:2005ub}. For
  recent reviews see
  Refs.~\cite{Trigiante:2007ki,Weidner:2006rp,Samtleben:2008pe,deWit:2009zv}. }
of general bosonic 4-dimensional field theories has recently been elucidated
in Ref.~\cite{Bergshoeff:2009ph} and applied to the search of higher-rank
$p$-form potentials in gauged $N=1,d=4$ supergravity in
Ref.~\cite{Hartong:2009az}.

It is natural to try to extend the recently obtained results on 4-dimensional
tensor hierarchies to higher dimensions. The 4-dimensional results suggest the
existence of some general features common to all d-dimensional tensor
hierarchies: 

\begin{enumerate}
\item The one-to-one relation between $(d-2)$-form potentials (which always carry an adjoint index) and the
  symmetries of the theory. We will henceforth refer to them as adjoint-form potentials or simply
  \textit{ad-form} potentials.

\item The one-to-one relation between the $(d-1)$-form potentials and the
  components of the embedding tensor (and, possibly, other \textit{deformation
    tensors}). Following Ref.~\cite{Bergshoeff:2007qi}, we will call these
  potentials \textit{de-form} potentials.

\item The one-to-one relation between the top- ($d$-) form potentials and all
  the constraints satisfied by the embedding tensor (and, possibly, other
  deformation tensors).
\end{enumerate}

\noindent
Some of these relations have been discussed in Ref.~\cite{deWit:2008gc}.

In this paper we are going to study in detail 5- and 6-dimensional field
theories and we are going to find the general rules that determine the
structure of their associated tensor hierarchies. The special case of maximal
supergravity in five and six dimensions has been considered in
Refs.~\cite{deWit:2004nw,Bergshoeff:2007ef}.

As we are going to see, there are important differences between the maximal
supergravity case and the general case, the principal difference being the
existence of more independent deformation tensors in addition to the embedding
tensor. These deformation tensors switch on new couplings such as massive
deformations, unrelated to (but compatible with) Yang-Mills gauge symmetries,
which are determined by the embedding tensor alone. In maximal supergravities,
supersymmetry determines these deformation tensors entirely in terms of the
gauge group and the embedding tensor. In the general case the deformation
tensors are, up to a few constraining relations, independent of the embedding
tensor.

Taking into account the existence of several deformation tensors we find that
the highest-rank potentials of the tensor hierarchy can be constructed as
follows. Let us denote by $A^{I}$ the 1-forms of the $d$-dimensional tensor
hierarchy, by $\vartheta_{I}{}^{A}$ the embedding tensor where $A$ is an
adjoint index of some symmetry group and by $c^{\sharp}$ the deformation
tensors (including the embedding tensor). Here $\sharp$ denotes the
corresponding indices. The magnetic duals of the 1-forms will be the
hierarchy's $(d-3)$-forms $\tilde{A}_{I}$, with $(d-2)$-form field strengths
$\tilde{F}_{I}$. These will contain a St\"uckelberg coupling to the
ad-form potentials that we are going to denote by $C_{A}$, and the
coupling tensor will be the embedding tensor $\vartheta_{I}{}^{A}$, so

\begin{equation}
\tilde{F}_{I}  \sim \mathfrak{D}\tilde{A}_{I}
+\cdots
+\vartheta_{I}{}^{A}C_{A}\, .
\end{equation}

The $(d-1)$-form field strength for $C_{A}$, denoted here by $G_{A}$, can be
obtained by hitting the above expression with a covariant derivative
$\mathfrak{D}$. This gives rise to an expression for $\vartheta_I{}^A G_A$ and
determines $G_A$ up to terms that vanish upon contraction with
$\vartheta_I{}^A$. These extra terms in $G_A$ form St\"uckelberg couplings to
de-form potentials. The coupling tensors will vanish upon contraction (of
the adjoint index) with the embedding tensor. They can be constructed in the
following way. All the deformation tensors must be gauge-invariant tensors,
and, if their gauge transformations are written as

\begin{equation}
\label{eq:constraintconvention}
\delta_{\Lambda} c^{\sharp} = -\Lambda^{I} Q_{I}{}^{\sharp}\, , 
\end{equation}

\noindent
where the $\Lambda^{I}(x)$ are the 0-form gauge transformation parameters of the
1-forms $A^{I}$, then, we find a constraint

\begin{equation}
 Q_{I}{}^{\sharp}\equiv -\delta_{I} c^{\sharp}=0\, ,  
\end{equation}

\noindent
for each of them. All these constraints are, by construction, proportional to
the embedding tensor

\begin{equation}
\delta_{\Lambda} c^{\sharp} = \Lambda^{I}\vartheta_{I}{}^{A}\delta_{A}c^{\sharp}\, ,  
\end{equation}

\noindent
and can be written in the form

\begin{equation}
 Q_{I}{}^{\sharp} = -\vartheta_{I}{}^{A}Y_{A}{}^{\sharp}\, ,  
\hspace{1cm}
Y_{A}{}^{\sharp} \equiv \delta_{A}c^{\sharp}\, ,
\end{equation}

\noindent
which provides us with as many tensors $Y_{A}{}^{\sharp}$ as we have deformation
tensors $c^{\sharp}$. We will follow the above convention to normalize
the constraints $Q$ and associated $Y$-tensors.

The $(d-1)$-form field strengths will have the form

\begin{equation}
G_{A}  \sim \mathfrak{D}C_{A}+\cdots
+\sum_{\sharp}Y_{A}{}^{\sharp}D_{\sharp}\, .
\end{equation}

\noindent
where we have introduced as many de-form potentials $D_{\sharp}$ as we have
deformation tensors $c^{\sharp}$, transforming in the representation conjugate
to the representation in which the $c^\sharp$ transform. This is precisely the
number of de-form potentials that we need to introduce in the action as
Lagrange multipliers enforcing the constancy of the deformation tensors

\begin{equation}
\int \sum_{\sharp} d c^{\sharp} \wedge D_{\sharp}\, .  
\end{equation}

Finally, the $d$-form field strengths $K_{\sharp}$ of the de-form potentials
$D_{\sharp}$ will have St\"uckelberg couplings to top-form potentials.  As
different from the 4-dimensional case in which there is only one $Y$-tensor
and the St\"uckelberg coupling tensors ($W$) are annihilated by the
$Y$-tensor, in the general case the $W$-tensors are not individually
annihilated by the $Y$-tensors. Instead, there are combinations of $Y$- and
$W$-tensors that vanish.

These combinations can be found systematically as follows. Let us introduce as
many top-form potentials as there are constraints satisfied by the deformation
tensors. This is precisely the number of top-forms that we need to introduce
in the action as Lagrange multipliers enforcing all the algebraic constraints.
We will have top forms $E^{I}{}_{\sharp}$ associated to the constraints
$Q_{I}{}^{\sharp}$ that express the gauge-invariance of the deformation
tensors, but we will have more top-forms, associated to other constraints. Let
us denote all the constraints satisfied by all the deformation tensors
$Q^{\flat}$ and the top forms by $E_{\flat}$ and let us construct the formal
combination

\begin{equation}
\sum_{\flat}Q^{\flat} E_{\flat}\, ,
\end{equation}

\noindent
which vanishes because it is linear in the constraints.  This is the term one
needs to add to the action in order to enforce the constraints $Q^{\flat}=0$.

The infinitesimal linear transformations of this term generated by the
matrices $T_{A}$, that we will denote by $\delta_{A}$, also vanish because
these transformations are proportional to the constraints $Q^{\flat}$.  Since
the constraints $Q^{\flat}$ are functions of the deformation tensors, using
the chain rule we can write this vanishing infinitesimal transformation as

\begin{equation}
  0=\delta_{A}\left(\sum_{\flat}Q^{\flat} E_{\flat}\right)
  = 
  \sum_{\flat}
\left(\sum_{\sharp} \delta_{A}c^{\sharp} \frac{\partial Q^{\flat}}{\partial c^{\sharp}}\right) E_{\flat}
  =
  \sum_{\flat}
\left(\sum_{\sharp} Y_{A}{}^{\sharp} \frac{\partial Q^{\flat}}{\partial c^{\sharp}}\right) E_{\flat}\, ,
\end{equation}

\noindent
where we have made use of the general definition of the $Y$-tensors
Eq.~(\ref{eq:constraintconvention}). Since, in this expression, the top forms
$E_{\flat}$ have arbitrary values, we get, for each of them, the identity 

\begin{equation}
\label{eq:YWidentity}
\sum_{\sharp} Y_{A}{}^{\sharp} W_{\sharp}{}^{\flat}=0\, ,
\end{equation}

\noindent
where we have defined the $W$-tensors

\begin{equation}\label{eq:Wtensor}
 W_{\sharp}{}^{\flat}
\equiv \frac{\partial Q^{\flat}}{\partial c^{\sharp}}\, . 
\end{equation}

Then, the $d$-form field strengths $K_{\sharp}$ of the de-form potentials
$D_{\sharp}$ will have the general form 

\begin{equation}
K_{\sharp} \sim \mathfrak{D} D_{\sharp}+\cdots +\sum_{\flat} W_{\sharp}{}^{\flat}
E_{\flat}\, .
\end{equation}

This scheme leads to a number of ad-form potentials $C_{A}$ equal to the
number of (continuous) symmetries and, therefore, to Noether current
1-forms $j_{A}$. This is what we expect since, in order
not to add further continuous degrees of freedom to the theory the
$(d-1)$-form field strengths $G_{A}$ must be dual to the Noether currents

\begin{equation}
G_{A}\sim \star j_{A}\, .  
\end{equation}

\noindent
This scheme also leads to a number of de-form potentials $D_{\sharp}$
that is equal to the number of deformation tensors $c^{\sharp}$. As mentioned
above, we need this number of deformation tensors to enforce the constraints
$dc^{\sharp}=0$ in the action. With a Lagrange multiplier term enforcing the
constancy of the deformation tensors we can also vary the action with respect
to the deformation tensors which have off-shell been promoted to fields. This
leads to duality relations for their $d$-form field strengths $K_{\sharp}$ of
the form

\begin{equation}
  K_{\sharp} \sim \star \frac{\partial V}{\partial c^{\sharp}}\, .  
\end{equation}

\noindent
Finally, as already said, this scheme leads to one top-form potential for each
constraint satisfied by the deformation tensors.

The tensor hierarchy can be considered to be a technique that can be used to
predict in which way a given theory can be deformed. To make such a prediction
one can construct the de- and top-form field content of a particular
theory. The above scheme is only based on necessary conditions and is not
guaranteed to be sufficient to construct all possible de- and top-form
potentials of a particular (bosonic) field theory\footnote{When there are also
  fermions the tensor hierarchy may get extended due to ad-forms that are
  dual to currents bilinear in fermions that appear in the 1-form equations of
  motion. These ad-forms may then have St\"uckelberg couplings with new
  de-forms, etc. This has been shown to happen in $N=1$, $d=4$
  supergravity in Ref.~\cite{Hartong:2009az}.}. In order to see in which
manner the above described construction of the de-forms is not sufficient
let us consider possible sources of it failing to be so. For example, it could
happen that in order for $G_A$ to transform gauge-covariantly we need to
introduce a St\"uckelberg coupling with a tensor $Y_A$ which is not of the
form $\delta_A c$ where $c$ is some deformation tensor but which nonetheless
satisfies $\vartheta_I{}^AY_A{}=0$. Even though we have never encountered such
a $Y$-tensor we have not been able to disprove their existence. Similarly,
there may be additional top-forms contracted with $W$-tensors that are not of
the form Eq.~(\ref{eq:Wtensor}), but which nonetheless satisfy
Eq.~(\ref{eq:YWidentity}). Once again we did not prove that every $W$-tensor
that satisfies Eq.~(\ref{eq:YWidentity}) is of the form Eq.~(\ref{eq:Wtensor})
but we are not aware of any counterexamples. Another source of failure of the
above described program to find all the de- and top-form potentials is
that there may exist de- and top-form potentials which cannot appear in
any St\"uckelberg couplings. This happens for example in $N=1$, $d=4$
supergravity where there exists a 3-form potential that is dual to the
superpotential Ref.~\cite{Hartong:2009az}. This 3-form does not show up in any
of the St\"uckelberg couplings of the 4-dimensional tensor hierarchy and there
exists no choice of deformations tensors for which it would show up in a
St\"uckelberg coupling.

The construction of any tensor hierarchy starts with writing down the most
general form of the 2-form field strength $F^I$ which includes both
Yang--Mills pieces as well as St\"uckelberg couplings to 2-forms. From this
field strength, which at this stage should be thought of as an Ansatz, one can
construct a Bianchi identity by hitting it with a covariant derivative
$\mathfrak{D}$. From $\mathfrak{D} F^I$ we can obtain that part of the field
strength of the 2-forms that does not contain the St\"uckelberg couplings to
the 3-forms. By making once again an Ansatz for such a coupling we can proceed
to compute the Bianchi identity of the 3-form field strengths and continue in
this way until we reach the $d$-form field strengths of the de-form
potentials which contain St\"uckelberg couplings to the top-form
potentials. The Ans\"atze made throughout this procedure will then lead to a
nested set of Bianchi identities provided the various St\"uckelberg coupling
tensors satisfy certain relations. Once these relations have been obtained we
have at our disposal the most general set of tensor couplings\footnote{As
  mentioned before the tensor hierarchy does not predict those potentials that
  cannot appear in the St\"uckelberg couplings. These tensors must be dealt
  with separately.}  that a particular bosonic theory can have and we may
proceed to construct Lagrangians for these tensors.

This program will be performed in detail in Section~\ref{sec-d5hierarchy} for
the case of 5-dimensional field theory and in the
Section~\ref{sec-d6hierarchy} for the case of 6-dimensional field theory.


\section{The $d=5$ general tensor hierarchy}
\label{sec-d5hierarchy}


\subsection{$d=5$ Bosonic field  theories}
\label{sec-d5bosonicgaugetheories}

In $d=5$ dimensions vectors are dual to 2-forms. We can, therefore, use as a
starting point, theories with spacetime metric $g_{\mu\nu}$, scalars
$\phi^{x}$ parametrizing a target space with metric $g_{xy}(\phi)$ and 1-forms
$A^{I}$ only. The most general action with (ungauged and massless) Abelian
gauge-invariance $\delta A^I=-d\Lambda^I$, no gauged symmetries and terms with
no more than two derivatives that we can write for these fields
is\footnote{Our conventions for differential forms, Hodge duals etc. can be
  found in Appendix~\ref{app-conventions}.}

\begin{equation}\label{eq:actionungauged}
S  =   {\displaystyle\int} 
\biggl\{
\star R
+{\textstyle\frac{1}{2}}g_{xy}(\phi)d\phi^{x}\wedge \star d\phi^{y}
-{\textstyle\frac{1}{2}} a_{IJ}(\phi) F^{I}\wedge \star F^{J}
-\star V(\phi)
+{\textstyle\frac{1}{3}}C_{IJK}F^{I}\wedge F^{J}\wedge A^{K}
\biggr\}\, ,
\end{equation}

\noindent
where

\begin{equation}
\label{eq:FIAbelian}
F^{I}=dA^{I}\, ,  
\end{equation}

\noindent
and where $g_{xy}(\phi)$ and $a_{IJ}(\phi)$ are symmetric, positive-definite
matrices that depend on the scalar fields, $V(\phi)$ is a scalar potential and
$C_{IJK}$ is a constant, totally symmetric, tensor; any other components of
$C_{IJK}$ apart from the totally symmetric ones would not contribute to the
action and, therefore, without loss of generality, they are set equal to zero.

This action takes exactly the same form as the bosonic action of minimal $d=5$
supergravity coupled to vector supermultiplets and hypermultiplets (if we
assume all the corresponding scalars are represented by the $\phi^{x}$) given
in Ref.~\cite{Bergshoeff:2004kh}. However, although probably most interesting
applications of this work will be in the context of supergravity theories, we
stress that here we are considering a general field theory in which there is
no underlying real special geometry, the objects $g_{xy}(\phi)$,
$a_{IJ}(\phi)$, and $C_{IJK}$ need not be related by real special geometry as
in the supersymmetric case and the scalars parametrize arbitrary target spaces
and occur in a number which is unrelated to the number of vector fields.

From this point of view, the tensor $C_{IJK}$ is just a set of possible
deformations of the minimally coupled theory (which has $C_{IJK}=0$). It gives
rise to vector couplings unrelated to Yang-Mills gauge symmetry. This type of
couplings are not possible in $d=4$ dimensions.

If we only vary the 1-forms in the action, we get

\begin{equation}
\label{eq:MaxwelleqsAbelian}
\delta S = \int 
\left\{-\delta A^{I}\wedge \star \frac{\delta S}{\delta A^{I}}\right\}  \, ,
\hspace{1cm}
\star\frac{\delta S}{\delta A^{I}}
=
d(a_{IJ}\star F^{J})- C_{IJK} 
F^{J}\wedge F^{K}\, ,  
\end{equation}

\noindent
and, on account of Eq.~(\ref{eq:FIAbelian}), the equation of motion can be
rewritten in the form

\begin{equation}
d(a_{IJ}\star F^{J}-C_{IJK} F^{J}\wedge A^{K})=0\, .  
\end{equation}

\noindent
This suggests to define the 2-forms $B_{I}$ dual to the 1-forms $A^{I}$ via

\begin{equation}
a_{IJ}\star F^{J}- C_{IJK} F^{J}\wedge A^{K} \equiv dB_{I}\, .
\end{equation}

\noindent
Since, by definition, $a_{IJ}\star F^{J}$ is gauge-invariant,
the gauge-invariant field strengths of the 2-forms can be defined by

\begin{equation}
H_{I}\equiv dB_{I} +C_{IJK} A^{J}\wedge dA^{K}\, ,
\end{equation}

\noindent
so that we have the Bianchi identity and duality relation
\begin{equation}
dH_{I} = C_{IJK} F^{J}\wedge F^{K}\, , \qquad H_{I}=a_{IJ}\star F^{J}\, .  
\end{equation}

The gauge transformations of the 1- and 2-forms can be inferred from the
gauge-invariance of their field strengths:

\begin{eqnarray}
\delta_{\Lambda}A^{I} & = & -d\Lambda^{I}\, ,  \\
& & \nonumber \\
\delta_{\Lambda}B_{I} & = & d\Lambda_{I}+C_{IJK}\Lambda^{J}F^{K}\, . 
\end{eqnarray}

The construction of the tensor hierarchy based on the embedding-tensor
formalism should reproduce these results in the ungauged limit
$\vartheta_{I}{}^{A}$ (with any possible other deformation tensor not being
$C_{IJK}$ sent to zero as well).


\subsection{Gaugings and massive deformations}
\label{sec-d5gauging}

Let us consider the infinitesimal global transformations with constant parameters
$\alpha^{A}$ of the scalars $\phi^{x}$, 1-forms $A^{I}$ and dual 2-forms
$B_{I}$:

\begin{eqnarray}
\delta_{\alpha}\phi^{x} & = & \alpha^{A}k_{A}{}^{x}(\phi)\, ,  \\
& & \nonumber \\
\delta_{\alpha}A^{I} & = & \alpha^{A}T_{A\, J}{}^{I}A^{J}\, , \\
& & \nonumber \\
\delta_{\alpha}B_{I} & = & -\alpha^{A}T_{A\, I}{}^{J}B_{J}\, ,
\end{eqnarray}

\noindent
where the matrices $T_{A}$ belong to some representation of a group $G$ and
the $k_{A}{}^{x}(\phi)$ are the contravariant components of vectors defined on
the scalar manifold. Some of the matrices and the vectors may be identically
zero. They satisfy the algebras

\begin{equation}
\label{eq:Tkalgebras}
[T_{A},T_{B}]= -f_{AB}{}^{C}T_{C}\, ,
\hspace{1cm}  
[k_{A},k_{B}]= -f_{AB}{}^{C}k_{C}\, .
\end{equation}

These transformations will be global symmetries of the theory constructed in
the previous section if the following four conditions are met:

\begin{enumerate}
\item The vectors $k_{A}{}^{x}(\phi)$ are Killing vectors of the metric
  $g_{xy}(\phi)$ of the scalar manifold.
\item The kinetic matrix $a_{IJ}$ satisfies the condition

\begin{equation}
\label{eq:kineticmatrix}
\pounds_{A}a_{IJ}= -2 T_{A\, (I}{}^{K}a_{J)K}\, ,  
\end{equation}

\noindent 
where $\pounds_A$ denotes the Lie derivative along the vector $k_A$.

\item The deformation tensor is invariant 

\begin{equation}
\label{eq:YAIJK}
\delta_{A}C_{IJK} \equiv  Y_{A\, IJK} = -3T_{A\, (I}{}^{L}C_{JK)L}=0\, .    
\end{equation}

\item The scalar potential is invariant

\begin{equation}
\label{eq:potential}
\pounds_{A}V= k_{A}V=0\, .    
\end{equation}

\end{enumerate}

In what follows, we will relax these conditions. Conditions 1 and 2 above
cannot be relaxed but it is unnecessarily restrictive to demand that the
symmetry group of the minimally coupled undeformed theory which has
$C_{IJK}=0$ and $V=0$ is equal to the symmetry group $G$. More generally we
can allow $\delta_{A}C_{IJK}=Y_{A\, IJK}\neq 0$ and $\pounds_{A}V= k_{A}V\neq
0$ and instead consider that subgroup of $G$ under which $C_{IJK}$ and $V$ are
invariant. In this way we have the situation that $C_{IJK}$ and $V$ introduce
deformations that break the symmetry group $G$ of the undeformed theory to a
subgroup of $G$.

From the point of view of the construction of gauge-invariant theories using
the embedding tensor formalism the above conditions 3 and 4 are also
unnecessary. In general, the embedding tensor projects the above
transformations into a smaller subgroup of $G$. The theory that we will
construct will be only required to be invariant under gauge transformations of
this smaller subgroup, but not necessarily under all the above global
transformations. In the ungauged limit, i.e. setting the embedding tensor
equal to zero, the theory will be invariant under the global transformations
of the gauge group and not necessarily under any other global transformations.

From the general construction of the de- and top-form potentials, explained in the introduction, we know that if the tensor $C_{IJK}$ is invariant under
the transformations generated by all the matrices $T_{A}$, then the tensor
$Y_{A\, IJK}$ will vanish identically and there will not be a non-trivial
4-form potential $D^{IJK}$ dual to $C_{IJK}$.  There are cases of physical
interest (such as the maximal $d=5$ supergravity of Ref.~\cite{deWit:2004nw})
in which this is what happens.

After these comments, we can now proceed to gauge the above transformations.
This can be done by promoting the constant parameters $\alpha^{A}$ to
arbitrary functions and using the 1-forms as gauge fields. The embedding
tensor $\vartheta_{I}{}^{A}$ will relate the symmetry to be gauged with the
1-form that will gauge it:

\begin{equation}
\alpha^{A}(x) \equiv \Lambda^{I}\vartheta_{I}{}^{A}\, .  
\end{equation}

Thus, we want the theory to be invariant under the local transformations of
the scalars

\begin{equation}
\delta_{\Lambda}\phi^{x} = \Lambda^{I}\vartheta_{I}{}^{A}k_{A}{}^{x}(\phi)\, ,  
\end{equation}

\noindent
and for this we need the covariant derivatives

\begin{equation}
\label{eq:covariantderivativescalars}
\mathfrak{D}\phi^{x} \equiv d\phi^{x}
+A^{I}\vartheta_{I}{}^{A}k_{A}{}^{x}(\phi)\, .  
\end{equation}

\noindent
It can be checked that $\mathfrak{D}\phi^{x}$ transforms covariantly if we
impose the quadratic constraint

\begin{equation}
\label{eq:QIJA}
Q_{IJ}{}^{A}
\equiv   
-\delta_{I}\vartheta_{J}{}^{A}
=
\vartheta_{I}{}^{B}T_{B\, J}{}^{K}\vartheta_{K}{}^{A} 
-\vartheta_{I}{}^{B}\vartheta_{J}{}^{C}f_{BC}{}^{A} =0\, ,
\end{equation}

\noindent
and impose that the vectors transform according to

\begin{equation}
\label{eq:gaugetransformation1-forms-1}
\delta_{\Lambda}A^{I} 
= 
-\mathfrak{D}\Lambda^{I} +\Delta A^{I}
= 
-(d\Lambda^{I}+\vartheta_{J}{}^{A}T_{A\, K}{}^{I}A^{J}\Lambda^{K})  
 +\Delta A^{I}\, , 
\hspace{1cm}
\vartheta_{I}{}^{A}\Delta A^{I} =0\, ,
\end{equation}

\noindent
where the term $\Delta A^{I}$ is, otherwise and so far, arbitrary.

The above quadratic constraint means that $\vartheta_{I}{}^{A}$ is an
invariant tensor since

\begin{equation}
  \delta_{\Lambda}\vartheta_{I}{}^{A}
  =-\Lambda^{J}Q_{JI}{}^{A} = \Lambda^{J}\vartheta_{J}{}^{B} Y_{B\, I}{}^{A}=0\, ,  
\end{equation}

\noindent
where

\begin{equation}
 Y_{A\, I}{}^{B} 
\equiv 
\delta_{A}\vartheta_{I}{}^{B}
=
\vartheta_{I}{}^{C}f_{AC}{}^{B}
-T_{A\, I}{}^{K}\vartheta_{K}{}^{B}\, , 
\end{equation}

\noindent
is the $Y$-tensor associated to the quadratic constraint according to the
general formalism explained in the introduction.


\subsubsection{The 2-form field strengths $F^{I}$}
\label{sec-d5FI}

The next step is to construct the field strength $F^{I}$ of the 1-forms. If we
take the covariant derivative of the scalars' covariant ``field strength''
$\mathfrak{D}\phi^{x}$ we find

\begin{equation}
\label{eq:Bianchiscalars}
\mathfrak{D}\mathfrak{D}\phi^{x}  = 
(dA^{I}+\tfrac{1}{2}X_{JK}{}^{I}A^{JK})\vartheta_{I}{}^{A}k_{A}{}^{x}\, ,
\end{equation}

\noindent
where, from now on, we use the shorthand notation\footnote{We will use a
  similar notation for exterior products of 2-forms and 3-forms throughout the rest
  of the paper, for example: $B_{IJ}\equiv B_{I}\wedge B_{J}$ etc.}

\begin{equation}
A^{I\cdots J} \equiv A^{I}\wedge \cdots  \wedge A^{J}\, ,
\hspace{1cm}
dA^{I\cdots J} \equiv dA^{I}\wedge \cdots  \wedge dA^{J}\, ,
\hspace{1cm}
F^{I\cdots J} \equiv F^{I}\wedge \cdots  \wedge F^{J}\, ,
\hspace{.3cm}
\mathrm{etc.}
\end{equation}

\noindent
and where we have defined, as is customary, the $X$ generators

\begin{equation}
X_{IJ}{}^{K} \equiv \vartheta_{I}{}^{A}T_{A\, J}{}^{K}\, .  
\end{equation}

Since the left hand side of the above Bianchi identity is covariant, by
construction, the right hand side is also covariant and it is
natural\footnote{Actually, it can be argued that this is the only solution
  that does not require the introduction of additional fields in the theory.}
to define

\begin{eqnarray}
\mathfrak{D}\mathfrak{D}\phi^{x}  
& = &  
F^{I}\vartheta_{I}{}^{A}k_{A}{}^{x}\, , 
\\
& & \nonumber \\
\label{eq:FI-1}
F^{I} 
& \equiv & 
dA^{I}
+\tfrac{1}{2}X_{JK}{}^{I}A^{JK}  +\Delta F^{I}\, ,
\\
& & \nonumber \\
\label{eq:deltaFI-1}
\vartheta_{I}{}^{A}\Delta F^{I} & = & 0\, .
\end{eqnarray}

\noindent
Requiring gauge-covariance of $F^{I}$ one finds that the term $\Delta F^{I}$
must transform according to

\begin{equation}
\label{eq:deltadeltaFI}
\delta_{\Lambda}\Delta F^{I}  = -\mathfrak{D}\Delta A^{I}
+2X_{(JK)}{}^{I}[\Lambda^{J}F^{K}+\tfrac{1}{2} A^{J}\wedge \delta_{\Lambda}A^{K}]
\, .
\end{equation}

In order to satisfy the constraint $\vartheta_{I}{}^{A}\Delta
F^{I}=\vartheta_{I}{}^{A}\Delta A^{I}=0$ we introduce a St\"uckelberg tensor
$Z^{IJ}$ satisfying

\begin{equation}
\label{eq:QAI}
Q^{AI}\equiv \vartheta_{J}{}^{A}Z^{JI}=0\, ,  
\end{equation}

\noindent
and define

\begin{equation}
\Delta F^{I} \equiv Z^{IJ}B_{J}\, ,
\hspace{1cm}
\Delta A^{I} \equiv -Z^{IJ}\Lambda_{J}\, ,
\end{equation}

\noindent
where $\Lambda_{I}$ are the 1-form gauge parameters under which the 2-forms
$B_{I}$ must transform. 

Observe that the constraint~(\ref{eq:QAI}) tells us that the 2-forms can only
occur as St\"uckelberg fields in the ungauged vector field strengths. Only the
ungauged vector fields can be eaten up by the 2-forms which will become
massive. We are thus describing through the introduction of $Z^{IJ}$ besides
gaugings also massive deformations of the theory described in
Section~(\ref{sec-d5bosonicgaugetheories}).

The gauge transformation of $\Delta F^{I}$ implies

\begin{equation}
Z^{IJ}\delta_{\Lambda}B_{J} = Z^{IJ} \mathfrak{D}\Lambda_{J} 
+2X_{(JK)}{}^{I}[\Lambda^{J}F^{K}+\tfrac{1}{2} A^{J}\wedge
\delta_{\Lambda}A^{K}]\, .
\end{equation}

\noindent
This solution will only work if $X_{(JK)}{}^{I}\sim Z^{IL} \mathcal{O}_{JKL}$
for some tensor $\mathcal{O}_{JKL}$ symmetric, at least, in the last two
indices. It is natural to identify this tensor with the fully symmetric tensor
$C_{IJK}$ that we know can occur in a Chern-Simons term in the action. This
identification allows us to recover the theory of
Section~(\ref{sec-d5bosonicgaugetheories}) in the
$\vartheta_{I}{}^{A},Z^{IJ}\rightarrow 0$ limit.

Thus, we impose the constraint\footnote{In $d=4$ dimensions there is
  a similar constraint which is linear in the embedding tensor. In $d=5$ the
  constraint has terms linear and of zeroth order in the embedding tensor.}

\begin{equation}
\label{eq:QJKI}
Q_{JK}{}^{I}\equiv X_{(JK)}{}^{I}- Z^{IL} C_{JKL}=0\, , 
\end{equation}

\noindent
and find that the field strength

\begin{equation}
F^{I}=dA^{I}+\tfrac{1}{2}X_{JK}{}^{I}A^{JK}  +Z^{IJ}B_{J}\, ,
\end{equation}

\noindent
transforms gauge-covariantly under the gauge transformations:

\begin{eqnarray}
\delta_{\Lambda}A^{I} 
& = &  
-\mathfrak{D}\Lambda^{I} -Z^{IJ}\Lambda_{J}\, ,\\
& & \nonumber \\
\delta_{\Lambda}B_{J} 
& = & 
 \mathfrak{D}\Lambda_{J} 
+2C_{JKL}(\Lambda^{K}F^{L} +\tfrac{1}{2}A^{K}\wedge
\delta_{\Lambda}A^{L} ) +\Delta B_{J}\, ,
\hspace{.7cm}
 Z^{IJ}\Delta B_{J}=0\, ,
\end{eqnarray}

\noindent
where the possible additional term $\Delta B_{J}$ will be determined by the
requirement of gauge-covariance of the 3-form field strength $H_{J}$.

The St\"uckelberg tensor $Z^{IJ}$ and the Chern--Simons tensor $C_{IJK}$ have to
be gauge-invariant tensors, which, following the convention in
Eq.~(\ref{eq:constraintconvention}), leads to the constraints

\begin{eqnarray}
\label{eq:QIJK}
Q_{L}{}^{IJ}
& \equiv & 
\delta_{L} Z^{IJ} 
= 
-(X_{LK}{}^{I}Z^{KJ}+X_{LK}{}^{J}Z^{IK}) =0\, ,
\\
& & \nonumber \\
\label{eq:QIJKL}
Q_{IJKL}
& \equiv & 
\delta_{I} C_{JKL}
=
3 X_{I(J}{}^{M}C_{KL)M}= 0\, ,  
\end{eqnarray}

\noindent
and to the $Y$-tensors

\begin{equation}
\label{eq:YAIJ}
Y_{A}{}^{IJ} 
 \equiv 
\delta_{A} Z^{IJ}
=
T_{A\, K}{}^{I}Z^{KJ}+T_{A\, K}{}^{J}Z^{IK}\, , 
\end{equation}

\noindent
and $Y_{A\, IJK}$ given in Eq.~(\ref{eq:YAIJK}), which are both annihilated by
the embedding tensor by virtue of the above constraints.


\subsubsection{The 3-form field strengths $H_{I}$}
\label{sec-d5HI}

The covariant derivative of the 2-form field strength $F^{I}$, after use
of the generalized Jacobi identities

\begin{equation}
X_{[JK}{}^{M}X_{L]M}{}^{I} = \tfrac{2}{3}Z^{IN}X_{[JK}{}^{M}C_{L]MN}\, ,  
\end{equation}

\noindent
is

\begin{equation}
\mathfrak{D}F^{I} 
=
Z^{IJ}[\mathfrak{D}B_{J}+C_{JKL}A^{K}\wedge dA^{L}
+\tfrac{1}{3}C_{JP[K}X_{ML]}{}^{P}A^{KML}]\, ,
\end{equation}

\noindent
which leads us to define the 3-form field strength

\begin{eqnarray}
\label{eq:BianchiFI}
\mathfrak{D}F^{I} 
& = & 
Z^{IJ} H_{J}\, ,
\\
& & \nonumber \\
H_{J} 
& \equiv & 
\mathfrak{D}B_{J}+C_{JKL}A^{K}\wedge dA^{L}
+\tfrac{1}{3} C_{JP[K}X_{ML]}{}^{P}A^{KML}  +\Delta
H_{J}\, ,
\\
& & \nonumber \\
Z^{IJ}\Delta H_{J} & = & 0\, ,
\end{eqnarray}

\noindent
where $\Delta H_{J}$ will be determined, together with $\Delta B_{J}$ by
requiring gauge-covariance of $H_{J}$. Instead of constructing gauge
transformations realizing gauge-covariance we construct a Bianchi identity for
$H_I$ in terms of gauge-covariant objects.

Let us first take the covariant derivative of both sides of the Bianchi
identity of $F^{I}$ Eq.~(\ref{eq:BianchiFI}). Using the Ricci identity

\begin{equation}
\mathfrak{D}\mathfrak{D}F^{I}= X_{JK}{}^{I}F^{JK}= 
Z^{IL}C_{LJK}F^{JK}\, ,
\end{equation}

\noindent
we find

\begin{equation}
Z^{IL}(\mathfrak{D}H_{L}-C_{LJK}F^{JK}) =0\, ,  
\end{equation}

\noindent
which implies that the Bianchi identity for $H_{I}$ must have the
form\footnote{$\Delta \mathfrak{D}H_{I}$ should not be confused with
  $\mathfrak{D}\Delta H_{I}$.}

\begin{equation}
\mathfrak{D}H_{I}=C_{IJK}F^{JK} +\Delta \mathfrak{D}H_{I}\, ,
\hspace{1cm}
Z^{JI}\Delta \mathfrak{D}H_{I}=0\, ,
\end{equation}

\noindent
which, in turn, implies that $\Delta \mathfrak{D}H_{I}$ must be proportional
to the invariant tensor(s) we mentioned before. To find them, we have to
compute directly $\mathfrak{D}H_{I}$ using the above expression.

In order to make progress in the calculation we must impose the  constraint

\begin{equation}\label{eq:ASofZ}
Z^{IJ}=-Z^{JI}\, .  
\end{equation}

\noindent
This property implies that the quadratic constraint $Q_{I}{}^{JK}$ and tensor
$Y_{A}{}^{JK}$ can be written in the form

\begin{equation}
\label{eq:QIJK2}
Q_{I}{}^{JK}=
2X_{IL}{}^{[J}Z^{K]L}\, ,
\hspace{1cm} 
Y_{A}{}^{JK}
= 
-2T_{A\, L}{}^{[J}Z^{K]L} \, .
\end{equation}

\noindent
A tensor with properties similar to those of $Z^{IJ}$ appears in $N=2$, $d=5$
supergravity with general couplings to vector and tensor supermultiplets in
Ref.~\cite{Bergshoeff:2004kh}.


\subsubsection{The 4-form field strengths $G_{A}$}
\label{sec-d5GA}

Using Eqs.~(\ref{eq:QAI}) and (\ref{eq:ASofZ}) we find that $\Delta H_{I}$ and
$\Delta\mathfrak{D}H_{I}$ can be taken to be

\begin{equation}
\Delta H_{I}= \vartheta_{I}{}^{A}C_{A}\, ,
\hspace{1cm}
\Delta \mathfrak{D}H_{I}= \vartheta_{I}{}^{A}G_{A}\, ,  
\end{equation}

\noindent
where $\vartheta_{I}{}^{A}G_{A}$ is the gauge-covariant field strength of the
3-forms $\vartheta_I{}^{A}C_{A}$. This determines the Bianchi identity of $H_{I}$ to
be

\begin{equation}
\label{eq:BianchiHI}
\mathfrak{D}H_{I}=C_{IJK}F^{JK}+\vartheta_{I}{}^{A}G_{A}\, .
\end{equation}

\noindent 
An explicit computation of $\mathfrak{D}H_{I}$ gives

\begin{eqnarray}
G_{A} 
& = & 
\mathfrak{D}C_{A} 
+T_{A\, K}{}^{I} 
\left[ 
  (F^{K}-\tfrac{1}{2}Z^{KL}B_{L})\wedge B_{I}
+\tfrac{1}{3}C_{ILM}A^{KL}\wedge dA^{M} 
\right.
\nonumber \\
& & \nonumber \\
& & 
\left.
+\tfrac{1}{12}C_{ILP}X_{MN}{}^{P}A^{KLMN} 
\right] +\Delta G_{A}\, ,
\\
& & \nonumber \\
\vartheta_{I}{}^{A} \Delta G_{A} & = & 0\, .
\end{eqnarray}

\noindent
According to the general scheme outlined in the introduction we expect that
$\Delta G_{A}$ will be formed out of terms proportional to the three
$Y$-tensors $Y_{A\,I}{}^{B}=\delta_{A}\vartheta_{I}{}^{B}$,
$Y_{A}{}^{IJ}=\delta_{A} Z^{IJ}$, $Y_{A\, IJK}=\delta_{A} C_{IJK}$ associated
to the three deformation tensors, contracted with some de-form
potentials. Each of these $Y$-tensors is annihilated by the embedding
tensor. We will next confirm that this is indeed what happens.


\subsubsection{The 5-form field strengths $K$}
\label{sec-d5K}

To find the invariant tensors and de-forms that make up $\Delta G_{A}$
we follow the same procedure as before and take the covariant derivative of
both sides of the Bianchi identity~(\ref{eq:BianchiHI}) for $H_{I}$. Using
the Ricci identity

\begin{equation}
\mathfrak{D}\mathfrak{D}H_{I}
= -\vartheta_{J}{}^{A} T_{A\, I}{}^{K}F^{J}\wedge H_{K}\, ,
\end{equation}

\noindent
and the Bianchi identities for $F^{I}$ and $H_{I}$, we get

\begin{equation}
\vartheta_{I}{}^{A}[\mathfrak{D}G_{A} -T_{A\, J}{}^{K}F^{J}\wedge H_{K}]=0\, ,  
\end{equation}

\noindent
from which it follows that the Bianchi identity for $G_{A}$ will have the form

\begin{equation}
\mathfrak{D}G_{A} =T_{A\, J}{}^{K}F^{J}\wedge H_{K} +\Delta
\mathfrak{D}G_{A}\, ,
\hspace{1cm}
\vartheta_{I}{}^{A}\Delta \mathfrak{D}G_{A}=0\, .
\end{equation}

\noindent
This implies that $\Delta \mathfrak{D}G_{A}$ must be proportional to the same
invariant tensors that $\Delta G_{A}$ is proportional to. A direct calculation
of $\mathfrak{D}G_{A}$ gives the result

\begin{equation}
  \begin{array}{rcl}
\mathfrak{D}G_{A}
& = & 
T_{A\, K}{}^{I}F^{K}\wedge H_{I}
\\
& & \\
& & 
+Y_{A}{}^{IJ}\left[\tfrac{1}{2}\mathfrak{D}B_{I}-H_{I}\right]\wedge B_{J} 
\\
& & \\
& & 
+Y_{A\, I}{}^{B}
\left[
(F^{I}-Z^{IL}B_{L})\wedge C_{B}
+\tfrac{1}{12}T_{B\, J}{}^{M}C_{KML}A^{IJK} \wedge dA^{L}
\right.
\\
& & \\
& & 
\left.
+\tfrac{1}{60}T_{B\, J}{}^{N}C_{KPN}X_{LM}{}^{P}A^{IJKLM}
\right]   
\\
& & \\
& & 
+Y_{A\, IJK}
\left[
\tfrac{1}{3}A^{I}\wedge dA^{JK}
+\tfrac{1}{4}X_{LM}{}^{K} A^{ILM}\wedge dA^{J}
+\tfrac{1}{20}X_{LM}{}^{J}X_{NP}{}^{K}A^{ILMNP}
\right]
\\
& & \\
& & 
+\mathfrak{D}\Delta G_{A}\, .
  \end{array}
\end{equation}

\noindent 
This tells us that we must introduce three de-forms $D^{IJ}$, $D^I{}_{A}$ and
$D^{IJK}$, with the same symmetries as the respective $Y$-tensors, and take

\begin{equation}
\Delta G_{A} =   
Y_{A}{}^{IJ}D_{IJ}+Y_{A\, I}{}^{B}D^I{}_{B}+Y_{A\, IJK}D^{IJK}\, ,
\end{equation}

\noindent 
in order for $\mathfrak{D}G_{A}$ to be gauge-covariant. This is simply because
the terms proportional to the $Y$-tensors must each be gauge-covariant and
this can only be the case of they form field strengths of de-forms. The
ad-form field strength $G_{A}$ and its Bianchi identity take the final form

\begin{eqnarray}
G_{A} 
& = & 
\mathfrak{D}C_{A} 
+T_{A\, K}{}^{I} 
\left[ 
  (F^{K}-\tfrac{1}{2}Z^{KL}B_{L})\wedge B_{I}
+\tfrac{1}{3}C_{ILM}A^{KL}\wedge dA^{M} 
+\tfrac{1}{12}C_{ILP}X_{MN}{}^{P}A^{KLMN} 
\right] 
\nonumber \\
& & \nonumber \\
& & 
+Y_{A}{}^{IJ}D_{IJ}+Y_{A\, I}{}^{B}D^I{}_{B}+Y_{A\, IJK}D^{IJK}\, ,
\\
& & \nonumber \\
\label{eq:BianchiGA}
\mathfrak{D}G_{A}
& = &   
T_{A\, K}{}^{I}F^{K}\wedge H_{I}
+Y_{A}{}^{IJ}K_{IJ}
+Y_{A\, I}{}^{B}K^I{}_{B}
+Y_{A\, IJK}K^{IJK}\, , 
\end{eqnarray}

\noindent
where

\begin{eqnarray}
K_{IJ}
& \equiv & 
\mathfrak{D} D_{IJ}
-\left[H_{[I}-\tfrac{1}{2}\mathfrak{D}B_{[I}\right]\wedge B_{J]}
+\Delta K_{IJ}\, , 
\\
& & \nonumber \\
K^I{}_{B}
& \equiv & 
\mathfrak{D} D^I{}_{B}
+(F^{I}-Z^{IL}B_{L})\wedge C_{B}
+\tfrac{1}{12}T_{B\, J}{}^{M}C_{KML}A^{IJK} \wedge dA^{L}
\nonumber \\
& & \nonumber \\
& & 
+\tfrac{1}{60}T_{B\, J}{}^{N}C_{KPN}X_{LM}{}^{P}A^{IJKLM}
+\Delta K^I{}_{B}\, ,
\\
& & \nonumber \\
K^{IJK}
& \equiv & 
\mathfrak{D} D^{IJK}
+\tfrac{1}{3}A^{(I}\wedge dA^{JK)}
+\tfrac{1}{4}X_{LM}{}^{(K} A^{I|LM}\wedge dA^{|J)}
+\tfrac{1}{20}X_{LM}{}^{(J}X_{NP}{}^{K}
A^{I)LMNP}
\nonumber \\
& & \nonumber \\
& & 
+\Delta K^{IJK}\, ,
\end{eqnarray}

\noindent
in which $\Delta K_{IJ}$, $\Delta K^{I}{}_{B}$ and $\Delta K^{IJK}$ satisfy

\begin{equation}
Y_{A}{}^{IJ}\Delta K_{IJ}
+
Y_{A\, I}{}^{B}\Delta K^{I}{}_{B}
+
Y_{A\, IJK} \Delta K^{IJK}  
=
0\, .
\end{equation}

As explained in the introduction the terms $\Delta K$ will be contractions of
($W$-)tensors and 5-form potentials. To determine the $W$-tensors and the
5-form potentials, we take the covariant derivative of the Bianchi identity of
$G_{A}$, Eq.~(\ref{eq:BianchiGA}). Ignoring the fact that we are working in
$d=5$ dimensions we get

\begin{equation}
Y_{A}{}^{IJ}[\mathfrak{D}K_{IJ} -\tfrac{1}{2}H_{IJ}]
+
Y_{A\, I}{}^{B}[\mathfrak{D} K^{I}{}_{B} -F^{I}\wedge G_{B}]
+
Y_{A\, IJK} [\mathfrak{D}K^{IJK}  -\tfrac{1}{3}F^{IJK}]
=
0\, .
\end{equation}

\noindent
If we take the covariant derivative of the above expression, we find

\begin{equation}
  \begin{array}{rcl}
F^{K}\wedge K_{MN} 
\{+2 Y_{A}{}^{IM}X_{KI}{}^{N} -Y_{A\, K}{}^{B}Y_{B}{}^{MN}
\}    
& & \\
& & \\
+F^{KL}\wedge H_{M}
\{
-Y_{A}{}^{IM}C_{KLI} -Y_{A\, L}{}^{B}T_{B\, K}{}^{M} -Y_{A\, IKL}Z^{IM}
\}
& & \\
& & \\
+G_{B}\wedge H_{J}
\{ 
-Y_{A}{}^{IJ}\vartheta_{I}{}^{B} -Y_{A\, I}{}^{B}Z^{IJ}
\}
& & \\
& & \\
+F^{I}\wedge K^{JKL}
\{
-Y_{A\, I}{}^{B}Y_{B\, JKL} +3 Y_{A\, MJK}X_{IL}{}^{M}
\}
& & \\
& & \\
+F^{K}\wedge K^{J}{}_{D}
\{
Y_{A\, I}{}^{B}W_{B}{}^{I}{}_{KJ}{}^{D}
\}
& = & 0\, ,\\ 
  \end{array}
\end{equation}

\noindent
where

\begin{equation}
W_{B}{}^{I}{}_{KJ}{}^{D}
\equiv
\vartheta_{K}{}^{C}f_{BC}{}^{D}  \delta_{J}{}^{I}
+X_{KJ}{}^{I}\delta_{B}{}^{D}
-Y_{B\, J}{}^{D}\delta_{K}{}^{I}\, ,
\end{equation}

\noindent 
as in $d=4$.

Each term in braces is linear (or quadratic) in $Y$-tensors and vanishes
identically upon use of the 5 constraints
$Q_{I}{}^{JK},Q_{IJ}{}^{K},Q^{A\,I},Q_{I\, JKL},Q_{IJ}{}^{A}$. Furthermore,
the index structure of the products of field strengths which multiply the 5
expressions in braces coincides with that of the duals of those 5
constraints. Actually, each of those terms corresponds to one of the
identities in Eq.~(\ref{eq:YWidentity}), and we can rewrite the above
expression in the form

\begin{equation}
  \begin{array}{rcl}
F^{I}\wedge K_{JK} 
\left\{Y_{A}{}^{LM}
{\displaystyle  \frac{\partial Q_{I}{}^{JK}}{\partial Z^{LM}}} 
+Y_{A\,L}{}^{B} 
{\displaystyle\frac{\partial Q_{I}{}^{JK}}{\partial \vartheta_{L}{}^{B}}} 
\right\}    
& & \\
& & \\
+F^{IJ}\wedge H_{K}
\left\{
Y_{A}{}^{LM}
{\displaystyle  \frac{\partial Q_{IJ}{}^{K}}{\partial Z^{LM}}} 
+Y_{A\, L}{}^{B}
{\displaystyle  \frac{\partial Q_{IJ}{}^{K}}{\partial \vartheta_{L}{}^{B}}} 
+Y_{A\, LMN}
{\displaystyle  \frac{\partial Q_{IJ}{}^{K}}{\partial C_{LMN}}} 
\right\}
& & \\
& & \\
+G_{B}\wedge H_{I}
\left\{ 
Y_{A}{}^{JK}
{\displaystyle  \frac{\partial Q^{B\, I}}{\partial Z^{JK}}} 
+Y_{A\, J}{}^{C}
{\displaystyle  \frac{\partial Q^{B\, I}}{\partial \vartheta_{J}{}^{C}}} 
\right\}
& & \\
& & \\
+F^{I}\wedge K^{JKL}
\left\{
Y_{A\, M}{}^{B}
{\displaystyle  \frac{\partial Q_{I\, JKL}}{\partial \vartheta_{M}{}^{B}}} 
+3 Y_{A\, MNP}
{\displaystyle  \frac{\partial Q_{I\, JKL}}{\partial C_{MNP}}} 
\right\}
& & \\
& & \\
+F^{I}\wedge K^J{}_{B}
\left\{
Y_{A\, K}{}^{C}
{\displaystyle  \frac{\partial Q_{IJ}{}^{B}}{\partial \vartheta_{K}{}^{C}}} 
\right\}
& = & 0\, .\\ 
  \end{array}
\end{equation}

The scheme explained in the introduction leads us to assume the existence of
five 5-forms $E^{I}{}_{JK},E^{IJ}{}_{K}, E_{A\, I}, E^{I\, JKL}, E^{IJ}{}_{A}$
dual to the 5 constraints $Q_{I}{}^{JK},Q_{IJ}{}^{K},Q^{A\,I},Q_{I\,
  JKL},Q_{IJ}{}^{A}$ so

\begin{eqnarray}
\Delta K_{IJ}
& \equiv & 
+2X_{K[I}{}^{L}E^{K}{}_{J]L}
-C_{KL[I}E_{J]}{}^{KL}
-\vartheta_{[I|}{}^{A}E_{A\, |J]}\, ,
\\
& & \nonumber \\
\Delta K_{B}{}^{I}
& \equiv &
W_{B}{}^{I}{}_{KJ}{}^{D} E^{KJ}{}_{D}
-Z^{IJ}E_{B\, J}
-T_{B\, K}{}^{J}E_{J}{}^{IK}
-Y_{B}{}^{JK}E^{I}{}_{JK}
\nonumber \\
& & \nonumber \\
& & 
-Y_{B\, JKM}E^{I\, JKM}\, ,
\\
& & \nonumber \\
\Delta K^{IJK}
& \equiv &
3 X_{LM}{}^{(I|}E^{L\, |JK)M} 
+Z^{L(I}E_{L}{}^{JK)}\, .
\end{eqnarray}

\noindent
Each of these expressions is of the form $\Delta K_{\sharp}= \sum_{\flat}
E_{\flat} \partial Q^{\flat}/\partial c^{\sharp} $.

With the determination of the 5-form field strengths $K$ we have completed the
construction of the 5-dimensional tensor hierarchy. The gauge transformations of all
the potentials can be obtained by constructing the most general gauge
transformations under which all the field strengths transform 
gauge-covariantly. We will not proceed to determine these gauge transformations as
they are in principle determined by the Bianchi identities.


\subsubsection{Gauge-invariant action for the 1- and 2-forms}
\label{sec-d5action}

The gauge-invariant action for the 1- and 2-forms is essentially the one
given in Ref.~\cite{deWit:2004nw}, with the $E_{6}$ tensors
$Z^{IJ},C_{IJK}$ replaced by arbitrary tensors satisfying the five
algebraic constraints, giving:

\begin{equation}
  \begin{array}{rcl}
S & = & 
{\displaystyle\int} 
\left\{
\star R
+{\textstyle\frac{1}{2}}g_{xy}(\phi)\mathfrak{D}\phi^{x}\wedge 
\star \mathfrak{D}\phi^{y}
-{\textstyle\frac{1}{2}} a_{IJ}(\phi) F^{I}\wedge \star F^{J}
-\star V(\phi)
\right.
\\
& & \\
& & 
-Z^{IJ}B_{I}\wedge [H_{J}-\tfrac{1}{2}\mathfrak{D}B_{J}]
+{\textstyle\frac{1}{3}}C_{IJK}
\left[
A^{I}\wedge dA^{JK}
+\tfrac{3}{4}X_{LM}{}^{I}A^{JLM} \wedge dA^{K}
\right.
\\
& & \\
& & 
\left.
\left.
+\tfrac{3}{20}X_{LM}{}^{I}X_{NP}{}^{J}
A^{LMNPK}
\right]
\right\}\, ,
\end{array}
\end{equation}

\noindent
where the scalar potential $V(\phi)$ may contain more terms than the one in
Eq.~(\ref{eq:actionungauged}). The new terms must depend on the deformation
tensors in such a way that the potential of the ungauged theory is recovered
when they are set to zero.

A general variation of the above action can be written in the
form\footnote{The tilde in the first variation w.r.t.~the 1-forms $A^{I}$
  defines a modified first variation which has a simpler form than the total
  first variation which would be, as usual, the sum of all the terms
  proportional to $\delta A^{I}$ and contains terms proportional to the
  equations of motion of other fields. We will use similar simplified first
  variations in the 6-dimensional action.}

\begin{equation}
\delta S \equiv \int 
\left\{
\delta g^{\mu\nu}  {\displaystyle\frac{\delta S}{\delta
g^{\mu\nu}}}
-\delta \phi^{x} \star \frac{\delta S}{\delta \phi^{x}}
-\delta A^{I}\wedge \star \frac{\widetilde{\delta S}}{\delta A^{I}}
-(\delta B_{I}-C_{IJK}A^{J}\wedge \delta A^{K}) \wedge 
\star \frac{\delta S}{\delta B_{I}}
\right\}  \, ,
\end{equation}

\noindent
where the equations of motion are\footnote{Explicitly, we have
  \begin{equation}
    \mathfrak{D}\star\mathfrak{D}\phi^{x}
=
    d\star\mathfrak{D}\phi^{x}
+\Gamma_{yz}{}^{x}\mathfrak{D}\phi^{y}\wedge \star \mathfrak{D}\phi^{z}
+\vartheta_{I}{}^{A}\partial_{y}k_{A}{}^{x} A^{I} \wedge \star
\mathfrak{D}\phi^{y}\, .
  \end{equation}
}

\begin{eqnarray}
{\displaystyle\frac{\delta S}{\delta g^{\mu\nu}}}
& = &
\star 
\left\{
G_{\mu\nu} 
+\tfrac{1}{2}g_{xy}[
\mathfrak{D}_{\mu}\phi^{x}\mathfrak{D}_{\nu}\phi^{y}
-\tfrac{1}{2}g_{\mu\nu}\mathfrak{D}_{\rho}\phi^{x}\mathfrak{D}^{\rho}\phi^{y}
]
\right.
\nonumber \\
& & \nonumber \\
& & 
\left.
-\tfrac{1}{2}
a_{IJ}[F^{I}{}_{\mu}{}^{\rho}F^{J}{}_{\nu\rho}
-\tfrac{1}{4}g_{\mu\nu}F^{I\, \rho\sigma}F^{J}{}_{\rho\sigma}]
+\tfrac{1}{2}g_{\mu\nu}V
\right\}
\, ,
\\
& & \nonumber \\
\star\frac{\delta S}{\delta \phi^{x}}
& = & 
g_{xy}\mathfrak{D}\star\mathfrak{D}\phi^{y}
+\tfrac{1}{2}\partial_{x}a_{IJ}F^{I}\wedge \star  F^{J}
+\star\partial_{x}V\, ,
\\
& & \nonumber \\
\star\frac{\widetilde{\delta S}}{\delta A^{I}}
& = & 
\mathfrak{D}(a_{IJ}\star F^{J}) 
-C_{IJK}F^{JK}
-\star \vartheta_{I}{}^{A}j_{A}\, ,
\\
& & \nonumber \\
\star \frac{\delta S}{\delta B_{I}}
& = & 
-Z^{IJ}(a_{JK}\star F^{K} -H_{J})\, ,
\end{eqnarray}

\noindent
in which we have defined the 1-form currents

\begin{equation}
j_{A} \equiv k_{A\, x}\mathfrak{D}\phi^{x}\, .  
\end{equation}

Now, we can substitute in the general variation of the action the gauge
transformations of the fields 

\begin{eqnarray}
\delta_{\Lambda} \phi^{x} 
& = & 
\Lambda^{I} \vartheta_{I}{}^{A}k_{A}{}^{x}\, ,\\
& & \nonumber \\
\delta_{\Lambda} A^{I}
& = &
-\mathfrak{D}\Lambda^{I} -Z^{IJ}\Lambda_{J}\, , \\
& & \nonumber \\
\delta_{\Lambda}B_{I}
& = & 
\mathfrak{D}\Lambda_{I} 
+2C_{IJK}(\Lambda^{J}F^{K} +\tfrac{1}{2}A^{J}\wedge
\delta_{\Lambda}A^{K} ) \, .
\end{eqnarray}

Checking invariance of the action under the gauge transformations generated by
0- and 1-form parameters amounts to checking the following two Noether
identities:

\begin{eqnarray}
\mathfrak{D}\star {\displaystyle \frac{\widetilde{\delta S}}{\delta A^{I}}}
+2C_{IJK}F^{J}\wedge \star  {\displaystyle \frac{\delta S}{\delta B_{K}}}
+\vartheta_{I}{}^{A} k_{A}{}^{x} 
\star{\displaystyle \frac{\delta S}{\delta \phi^{x}}}
& = & 0\, ,\\
& & \nonumber \\
\mathfrak{D}\star  {\displaystyle \frac{\delta S}{\delta B_{I}}}
+Z^{IJ}\star {\displaystyle \frac{\widetilde{\delta S}}{\delta A^{J}}}
& = & 0\, .
\end{eqnarray}

The second identity is easily seen to be satisfied. The first identity can
also be shown to be satisfied upon use of the Killing property of
$\vartheta_{I}{}^{A}k_{A}{}^{x}$, the property

\begin{equation}
\label{eq:kineticmatrix2}
\vartheta_{I}{}^{A}k_{A}a_{JK}
= -2 X_{I (J}{}^{L}a_{K)L}\, ,  
\end{equation}

\noindent
of the kinetic matrix, the condition

\begin{equation}
\label{eq:potential2}
\vartheta_{I}{}^{A}k_{A}V 
= 
0\, ,
\end{equation}

\noindent
of the scalar potential and the constraint $Q_{I\, JKL}=0$. Observe that these
are the same conditions required by global invariance but projected with the
embedding tensor, which means they are weaker conditions.

We can now relate the equations of motion derived from this action and the
tensor hierarchy's Bianchi identities via the duality relations

\begin{eqnarray}
a_{IJ}\star F^{J} 
& = & 
H_{I}\, ,
\\
& & \nonumber \\
\star j_{A}
& = & 
G_{A}\, ,
\\
& & \nonumber \\
 \star \frac{\partial V}{\partial c^{\sharp}}
& = & 
K_{\sharp}\, .
\end{eqnarray}

With these duality relations, the 1-form equations of motion become the
Bianchi identities for the hierarchy's 3-form field strengths $H_{I}$.  The
projected scalar equations of motion $k_{A}{}^{x} \star{\displaystyle
  \frac{\delta S}{\delta \phi^{x}}}$ become the Bianchi identity of the
hierarchy's 4-form field strengths $G_{A}$. In order to show this one must use
the Killing property of the $k_{A}{}^{x}$, Eq.~(\ref{eq:kineticmatrix}) for
the kinetic matrix, and the following expression for $k_AV$

\begin{equation}
\label{eq:potential3}
k_{A}V 
= 
\sum_{\sharp}Y_{A}{}^{\sharp}\frac{\partial V}{\partial c^{\sharp}}\, .
\end{equation}

Now that we have completed the construction of the 5-dimensional tensor
hierarchy and provided an interpretation of the various potentials we
summarize these results in Table~\ref{table:5DTH}. We will explain the meaning of the table by discussing in detail the case of the 2-forms. The other forms then go analogously. 

We have seen 2-forms appearing in the field strengths of the 1-forms. These are ungauged 1-forms because the field strengths of the gauged 1-forms do not contain any 2-forms. These 2-forms are $Z^{IJ}B_J$. Their gauge transformations are of the form $Z^{IJ}\delta B_J=Z^{IJ}\mathfrak{D}\Lambda_J$ plus terms involving the 0-form gauge transformation parameter $\Lambda^I$, but not the 2-form gauge transformation parameter $\Lambda_A$. Therefore, all the gauge transformations that the $Z^{IJ}B_J$ have are massless gauge transformations. This is indicated in Table~\ref{table:5DTH} by the term ``massless'' in the column called ``gauge transformations''. Since the $Z^{IJ}B_J$ 2-forms appear in the field strength of the ungauged 1-forms they form St\"uckelberg pairs with these ungauged 1-forms. This is indicated in Table~\ref{table:5DTH} by ``ungauged $A^I$'' in the column ``St\"uckelberg pair with''. It is not possible to say, unless we explicitly know all the components of $Z^{IJ}$ exactly which 2-form $B_I$ forms a St\"uckelberg pair with which 1-form $A^I$. Further, we also indicated that the 2-forms $Z^{IJ}B_J$ whose field strengths are $Z^{IJ}H_J$ are dual to $Z^{IJ}a_{JK}F^K$ and that 2-forms with these gauge transformation properties can only exist whenever $Z^{IJ}\neq 0$. Besides the 2-forms $Z^{IJ}B_J$ there are also those which do not appear in the field strengths of the 1-forms. Such 2-forms fall into two categories depending on their gauge transformation properties. The first possibility is that their field strengths contain St\"uckelberg couplings to 3-forms. These exist for those $I$ for which the St\"uckelberg coupling tensor $\vartheta_I{}^A\neq 0$ and they will have massive gauge transformations.These 2-forms cannot also belong the $Z^{IJ}B_J$ type discussed earlier. Finally it can also happen that there are $I$ values for which the 2-forms are not forming any St\"uckelberg pair with either 1-forms or 3-forms. Such 2-forms occur for example in the theory in which there is no embedding tensor nor the St\"uckelberg tensor $Z$. More generally they can occur in the gauged theory but only for those $I$ for which $\vartheta_I{}^A=Z^{IJ}=0$. The other entries of Table~\ref{table:5DTH} should be read in an analogous fashion. 

The 1-forms have been left out from the table since they behave the same in any tensor hierarchy in any dimension.  There are always three types: 1). gauged 1-forms which always have massless gauge transformations and exist for all those $A$ for which $\vartheta_I{}^A\neq 0$, 2). ungauged 1-forms with massive gauge transformations which exist for all those $I$ for which $Z^{IJ}\neq 0$ and 3). ungauged 1-forms with massless gauge transformations which exist for all those $I$ for which $\vartheta_I{}^A=Z^{IJ}=0$.

We end the discussion of the 5-dimensional tensor hierarchy with some comments
about possible redundancy of potentials. Potentials that have massive gauge
transformations can be totally gauged away, but which particular potentials
have a massive gauge transformation (i.e.~which $p$-form potentials are
St\"uckelberg fields for a $(p+1)$-form potential) depends on the
St\"uckelberg tensors occurring in their field strengths, as shown in
Table~\ref{table:5DTH}. Using a massive gauge transformation with a $p$-form
(local) parameter to eliminate a $p$-form St\"uckelberg potential partially
fixes the standard (massless) gauge transformations of the associated
$(p+1)$-form potentials, which become massive. The top-forms are special
because they have massive gauge transformations but they are not St\"uckelberg
fields for any higher-rank potential.

For the $p$-forms with $p=1,2,3$ this would lead to a (partial) gauge fixing
of the 2-, 3- and 4-form gauge transformations. When this is done one can for
example eliminate some of the 3-forms $C_A$ for certain values of $A$. In the
case of the 4-forms it can happen, depending on the details, that an entire
form $D_\sharp$ can be gauged away. The 4-form massive gauge transformations
are of the form $\delta D_\sharp=-W_\sharp{}^\flat\Lambda_\flat$ where
$\Lambda_\flat$ is the 5-form gauge transformation parameter, $\delta
E_\flat=\mathfrak{D}\Lambda_\flat$. The massive gauge transformations of the
4-forms $\delta D_\sharp=-W_\sharp{}^\flat\Lambda_\flat$ can sometimes be used
to eliminate entirely some of the 4-forms $D_\sharp$. This happens for example
in gauged maximal supergravity where there is only one deformation tensor, the
embedding tensor, and hence there is only one 4-form. Similar statements apply
to the 5-forms $E_\flat$ that always come contracted with $W_\sharp{}^\flat$
and are thus determined up to massive gauge transformations of the type
$\delta E_\flat=\Sigma_\flat$ with $W_\sharp{}^\flat\Sigma_\flat=0$.

\begin{table}[h!]
\begin{center}
\begin{tabular}{|c|c|c|c|c|}
\hline
 &&&&\\
Potential & Gauge  & Interpretation & St\"uckelberg & Existence \\
& transformation & (field strength dual to) & pair with & \\
&&&&\\
\hline
&&&&\\
$B_I$ & massive & $a_{IJ}F^J$ & $\vartheta_I{}^AC_A$ & $\forall I\;:\;\vartheta_I{}^A\neq 0$\\
&&&&\\
\hline
&&&&\\
$Z^{IJ}B_J$ & massless & $Z^{IJ}a_{JK}F^K$ & ungauged $A^I$ & $\forall I\;:\; Z^{IJ}\neq 0$\\
&&&&\\
\hline
&&&&\\
$B_I$ & massless & $a_{IJ}F^J$ & none & $\forall I\;:\;\vartheta_I{}^A=Z^{IJ}=0$\\
&&&&\\
\hline
&&&&\\
$C_A$ & massive & current $j_A$ of & $Y_A{}^\sharp D_\sharp$ & $\forall A\;:\;Y_A{}^\sharp\neq 0$\\
&& symmetry broken by $V$ &&\\
&&&&\\
\hline
&&&&\\
$\vartheta_I{}^AC_A$ & massless & current $j_A$ of & $B_I$ & $\forall I\;:\;\vartheta_I{}^A\neq 0$\\
&& gauged symmetry &&\\
&&&&\\
\hline
&&&&\\
$C_A$ & massless & current $j_A$ of & none & $\forall A\;:\;Y_A{}^\sharp=\vartheta_I{}^A=0$\\
&& global symmetry &&\\
&&&&\\
\hline
&&&&\\
$D_\sharp$ & massive & $\partial V/\partial c^{\sharp}$ & $W_\sharp{}^\flat E_\flat$ & $\forall\sharp\;:\;W_\sharp{}^\flat\neq 0$\\
&&&&\\
\hline
&&&&\\
$Y_A{}^\sharp D_\sharp$ & massless & $Y_A{}^{\sharp}\partial V/\partial c^{\sharp}$ & $C_A$ & $\forall A\;:\;Y_A{}^\sharp\neq 0$\\
&&&&\\
\hline
&&&&\\
$D_\sharp$ & massless & $\partial V/\partial c^\sharp$ & none & $\forall\sharp\;:\;W_\sharp{}^\flat=Y_A{}^\sharp=0$\\
&&&&\\
\hline
&&&&\\
$W_\sharp{}^\flat E_\flat$ & massless & enforces constraints & $D_\sharp$ & $\forall\sharp\;:\;W_\sharp{}^\flat\neq 0$\\
&&&&\\
\hline
 \end{tabular}
 \caption{All the $p\ge 2$ forms of the 5-dimensional tensor hierarchy, their St\"uckelberg properties and physical interpretation.}
\label{table:5DTH}
\end{center}
\end{table}

\newpage

\section{The $d=6$ general tensor hierarchy}
\label{sec-d6hierarchy}


\subsection{$d=6$ Bosonic field  theories}
\label{sec-d6bosonicgaugetheories}

In $d=6$ dimensions we can have, apart from a spacetime metric and scalars
$\phi^{x}$, $n_{1}$ 1-forms $A^{i}$ and $n_{2}$ electric 2-forms $B^{\Lambda}$.
The 1-forms $A^i$ are dual to 3-forms $C_{i}$ and the electric
2-forms $B^\Lambda$ are dual to magnetic 2-forms $B_{\Lambda}$ (we will study their
definitions later). Furthermore, in $d=6$ dimensions we can have real (anti-)
self-dual 3-forms and, therefore, we can constrain the 2-forms to have (anti-)
self-dual 3-form field strengths. 

We will write down an action ignoring momentarily the (anti-)
self-duality constraint and impose it on the equations of motion derived
from that action, as it was done in $N=2B$, $d=10$ supergravity in
Refs.~\cite{Bergshoeff:1995sq,Bergshoeff:2001pv}. This can only be done
consistently if the field strengths and action are such that the Bianchi
identities transform into the equations of motion and viceversa under
electric-magnetic duality transformations of the 2-forms. In particular, if
the action has Chern-Simons terms of the form $H\wedge F\wedge A$ which give
rise to terms proportional to $F\wedge F$ in the equations of motion of the 2-forms, the
field strengths $H$ must necessarily have terms of the form $F\wedge
A$.
 
Taking into account, thus, the possibility of having (anti-) self-dual
2-forms, the most general action with (ungauged and massless) Abelian 
gauge-invariance, with no more than two derivatives that we can write for scalars,
vectors and (electric) 2-forms is, in differential form language\footnote{See
  Appendix~\ref{app-conventions}.},

\begin{equation}
\begin{array}{rcl}
S 
& = & 
{\displaystyle\int} 
\left\{
-\star R 
+ \tfrac{1}{2}g_{xy}(\phi) d\phi^{x}\wedge \star d\phi^{y}
- \tfrac{1}{2} a_{ij}(\phi)F^{i}\wedge \star F^{j}
\right.
\\
& & \\
& & 
\left.
+ \tfrac{1}{2} b_{\Lambda\Sigma}(\phi) H^{\Lambda}\wedge \star H^{\Sigma}
+ \tfrac{1}{2} c_{\Lambda\Sigma}(\phi) H^{\Lambda}\wedge H^{\Sigma}
+\star V(\phi)
+\varepsilon d_{\Lambda\,  ij}H^{\Lambda}\wedge F^{i}\wedge A^{j}
\right\}\, .    
\end{array}
\end{equation}

\noindent
In this expression, $F^{i}$ and $H^{\Lambda}$ are the 2- and 3-form field
strengths, defined by 

\begin{eqnarray}
F^{i}  
& \equiv & 
dA^{i}\, ,\\
& & \nonumber \\
H^{\Lambda}  
& \equiv & 
dB^{\Lambda} +d^{\Lambda}{}_{ij}A^{i}\wedge dA^{j}\, ,
\end{eqnarray}

\noindent
invariant under the Abelian gauge transformations

\begin{eqnarray}
\delta A^{i}  
& = & 
-d\Lambda^{i}\, ,\\
& & \nonumber \\
\delta B^{\Lambda}  
& = & 
d\Lambda^{\Lambda} +d^{\Lambda}{}_{ij}\Lambda^{i} dA^{j} \, .
\end{eqnarray}

\noindent
The scalar-dependent kinetic matrices
$g_{xy}(\phi),b_{\Lambda\Sigma}(\phi),a_{ij}(\phi)$ are symmetric. The first
two of them are positive-definite and the third is negative-definite. The
tensor $c_{\Lambda\Sigma}(\phi)$ is antisymmetric.  The constant tensors
$d_{\Lambda\,ij}$ and $d^\Lambda{}_{ij}$ have the symmetries\footnote{The
  Chern--Simons term containing $d_{\Lambda\, ij}$ in the Lagrangian is
  clearly symmetric in $ij$ up to total derivatives. The terms containing
  $d^{\Lambda}{}_{ij}$, which appear in the field strengths $H^\Lambda$ are
  symmetric up to a field redefinition of $B^{\Lambda}$.}

\begin{equation}
d_{\Lambda\,  ij} = d_{\Lambda\,  ji}\, ,
\hspace{1cm}
d^{\Lambda}{}_{ij} = d^{\Lambda}{}_{ji}\, ,
\end{equation}

\noindent
ans satisfy the constraint

\begin{equation}
\label{eq:dd}
d_{\Lambda\,  i(j}d^{\Lambda}{}_{kl)}=0 \, ,
\end{equation}

\noindent
for the last term in the action to be gauge-invariant. We will later choose
the arbitrary constant $\varepsilon$ to have simple duality rules for the
2-forms.

If we vary the 1-forms and 2-forms in the action, we get

\begin{equation}
\delta S = \int 
\left\{-\delta A^{i}\wedge \star \frac{\widetilde{\delta S}}{\delta A^{i}}
-(\delta B^{\Lambda}+d^{\Lambda}{}_{ij} A^{i}\wedge \delta A^{j})\wedge 
\star \frac{\delta S}{\delta B^{\Lambda}} \right\}  \, ,
\end{equation}

\noindent
where

\begin{eqnarray}
 \star \frac{\widetilde{\delta S}}{\delta A^{i}} 
& = & 
d\{ a_{ij}\star F^{j} -2 d^{\Lambda}{}_{ij}A^{j}\wedge 
[J_{\Lambda} +\varepsilon d_{\Lambda\, kl} A^{k}\wedge dA^{l}]
\nonumber \\
& & \nonumber \\
& & 
-2\varepsilon d_{\Lambda\, ij}H^{\Lambda}\wedge A^{j} -\tfrac{2}{3}\varepsilon 
d_{\Lambda\, ij} d^{\Lambda}{}_{kl}A^{jk}\wedge dA^{l}\}\, , 
\\
& & \nonumber \\
\star \frac{\delta S}{\delta B^{\Lambda}}  
& = & 
d \{  J_{\Lambda} +\varepsilon d_{\Lambda\, ij} A^{i}\wedge dA^{j} \}\, ,
\end{eqnarray}

\noindent
where we have defined

\begin{equation}
\label{eq:JL}
J_{\Lambda} \equiv b_{\Lambda\Sigma}\star H^{\Sigma} +c_{\Lambda\Sigma}H^{\Sigma}\, ,  
\end{equation}

\noindent
and where we have used the Bianchi identities and the property
Eq.~(\ref{eq:dd}) in order to write the equations of motion of the vector
fields as total derivatives.


\subsubsection{The magnetic 2-forms $B_{\Lambda}$}
\label{sec-d6bosonicgaugetheories2-forms}

The equations of motion of the 2-forms $B^\Lambda$ suggest the definition of
the magnetic 2-forms $B_{\Lambda}$ through

\begin{equation}
dB_{\Lambda} 
\equiv 
J_{\Lambda} +\varepsilon d_{\Lambda\, ij} A^{i}\wedge d A^{j}\, .   
\end{equation}

\noindent 
Since $J_{\Lambda}$ is gauge-invariant, we define the
dual 3-form field strengths by

\begin{equation}
H_{\Lambda} \equiv J_{\Lambda}= dB_{\Lambda} -\varepsilon d_{\Lambda\, ij} A^{i}\wedge d A^{j}\, .     
\end{equation}

\noindent
We set $\varepsilon=-1$ to make the magnetic and electric 3-form field
strengths as similar as possible. Thus, we can
replace the equations of motion of the electric 2-forms, via the above definition
of the magnetic field strengths, by a Bianchi identity.  

In $d=6$ dimensions it is possible to constrain the 2-forms to have self- or
anti-self-dual field strengths. We can write these constraints in the form

\begin{equation}\label{eq:selfduality}
\zeta_{\Lambda\Omega} (H^{\Omega} -\zeta^{\Omega\Sigma}J_{\Sigma})=0\, ,  
\end{equation}

\noindent
where $\zeta^{\Lambda\Sigma}=\zeta_{\Lambda\Sigma}$ is a diagonal matrix whose
diagonal components can only be $+1$ for self-dual 3-form field strengths,
$-1$ for anti-self-dual 3-form field strengths or $0$ for unconstrained 3-form
field strengths.  The (anti-)self-duality constraints will be consistent if
the Bianchi identity for $H^{\Lambda}$ becomes the equation of motion of
$B^{\Lambda}$ upon their use. The Bianchi identities and the equations of
motion are

\begin{eqnarray}
dH^{\Lambda} 
& = & 
d^{\Lambda}{}_{ij}F^i\wedge F^j\, ,  
\\
& & \nonumber \\
dJ_{\Lambda} 
& = & 
d_{\Lambda\, ij}F^i\wedge F^j\, .  
\end{eqnarray}

\noindent
By hitting Eq.~(\ref{eq:selfduality}) with an exterior derivative we find that the tensors $d^{\Lambda}{}_{ij}$, and $d_{\Lambda\, ij}$ must satisfy the constraint

\begin{equation}\label{eq:zetaconstraint1}
\zeta_{\Omega\Lambda}( d^{\Lambda}{}_{ij}-\zeta^{\Lambda\Sigma}d_{\Sigma\, ij})=0\, ,  
\end{equation}

\noindent
for consistency.


\subsubsection{The 3-forms $C_{i}$}
\label{sec-d6bosonicgaugetheories3-forms}

The form of the equations of motion of the 1-forms also suggests the
definition

\begin{eqnarray}
dC_{i} & \equiv & 
a_{ij}\star F^{j} -2 d^{\Lambda}{}_{ij}A^{j}\wedge 
[J_{\Lambda} -d_{\Lambda\, kl} A^{k}\wedge dA^{l}]
+2d_{\Lambda\, ij}H^{\Lambda}\wedge A^{j}
\nonumber \\
& & \nonumber \\
& & 
 +\tfrac{2}{3}d_{\Lambda\, ij}
d^{\Lambda}{}_{kl} A^{jk}\wedge dA^{l}\, ,  
\end{eqnarray}

\noindent
or, using the magnetic 2-forms  and the constraint Eq.~(\ref{eq:dd})

\begin{equation}
dC_{i} =
a_{ij}\star F^{j} -2 d^{M}{}_{ij}
[A^{j} \wedge dB_{M}+\tfrac{1}{3}d_{M\, kl} A^{jk}\wedge dA^{l}]\, ,  
\end{equation}

\noindent
where we have defined the $2n_{2}$-component vectors

\begin{equation}
\label{eq:so2n2n2vectors}
(B^{M}) 
\equiv   
\left(
  \begin{array}{c}
B^{\Lambda} \\ \\ B_{\Lambda} \\   
  \end{array}
\right)\, ,
\hspace{.5cm}
(d^{M}{}_{ij}) 
\equiv   
\left(
\begin{array}{c}
d^{\Lambda}{}_{ij} \\  d_{\Lambda\, ij} \\   
  \end{array}
\right)\, ,
\hspace{.5cm}
(d_{M\, ij}) 
\equiv   
\left(
 d_{\Lambda\, ij}\, ,\,\,\,\, d^{\Lambda}{}_{ij} 
\right)\, .
\end{equation}

The gauge-invariant 4-form field strengths $G_{i}$ can be defined as 

\begin{equation}
G_{i} \equiv dC_{i} +2 d_{M\, ij}[ A^{j}\wedge dB^{M}
+\tfrac{1}{3}d^{M}{}_{kl} A^{jk}\wedge dA^{l}]\, ,    
\end{equation}

\noindent
which is related to the 2-form field strengths by the duality relation

\begin{equation}
G_{i}=  a_{ij}\star F^{j}\, .
\end{equation}

The 3-forms $C_{i}$ can be redefined in order to make contact with the 3-forms
that appear naturally in the tensor hierarchy. The redefinition is

\begin{equation}\label{eq:redefinedGi}
C^{\rm old}_{i} 
 \longrightarrow  
C^{\rm new}_{i}+2d_{M\, ij}B^{M}\wedge A^{j}\, ,  
\end{equation}

\noindent
so that

\begin{equation}
G_{i}=dC^{\rm new}_{i} +2 d^{M}{}_{ij}[ dA^{j}\wedge B_{M}
+\tfrac{1}{3}d_{M\, kl} A^{jk}\wedge dA^{l}]\, .
\end{equation}
 
\noindent 
The Bianchi identity satisfied by $G_i$ is

\begin{equation}
 dG_i=2d^M{}_{ij}F^j\wedge H_M\,.
\end{equation}

\noindent
In order to derive this it is useful to note that Eq.~(\ref{eq:dd}) can also
be written as

\begin{equation}
\label{eq:ddsonn}
d_{M\, i(j} d^{M}{}_{kl)}=0\, . 
\end{equation}


\subsubsection{Symmetries}
\label{sec-d6bosonicgaugetheoriessymmetries}

Let us momentarily set the $d$- and $\zeta$-tensors to zero and consider the
symmetries of the system of equations of motion and Bianchi identities of the
2-forms:

\begin{eqnarray}
d H^{\Lambda}   & = & 0\, ,\\
& & \nonumber \\
d J_{\Lambda}   & = & 0\, .
\end{eqnarray}

\noindent
This system is formally invariant under the $GL(2n_{2},\mathbb{R})$
transformations 

\begin{equation}
J^{M\prime} = M_{N}{}^{M}J^{N}\, ,
\hspace{.5cm}
(J^{M}) \equiv 
\left(
  \begin{array}{c}
H^{\Lambda} \\ \\ J_{\Lambda} \\   
  \end{array}
\right)\, .
\end{equation}

\noindent
These transformations must be consistent with the definition of $J_{\Lambda}$ 
in terms of $H^{\Lambda}$. Writing

\begin{equation}
(M_{N}{}^{M} )
\equiv 
\left(
  \begin{array}{cc}
A_{\Sigma}{}^{\Lambda} & B^{\Sigma\Lambda} \\
& \\
C_{\Sigma\Lambda} & D^{\Sigma}{}_{\Lambda} \\
\end{array}
\right)\, ,
\end{equation}

\noindent
we find that, for consistency, the symmetric and antisymmetric kinetic
matrices $b_{\Lambda\Sigma},c_{\Lambda\Sigma}$ must transform according to 

\begin{eqnarray}
f^{\prime} & = & (C+Df)(A+Bf)^{-1}\, ,\\
& & \nonumber   \\
f^{T\prime} & = & -(C-Df^{T})(A-Bf^{T})^{-1}\, ,
\end{eqnarray}

\noindent
where we have defined the matrix

\begin{equation}
f_{\Lambda\Sigma}=  b_{\Lambda\Sigma}+c_{\Lambda\Sigma}\, .
\end{equation}

\noindent
Consistency between the two transformation rules implies 

\begin{equation}
A^{T}C+C^{T}A=0\, ,
\hspace{1cm}  
B^{T}D+D^{T}B=0\, ,
\hspace{1cm}  
A^{T}D+C^{T}B=\xi \mathbb{I}_{n_{2}\times n_{2}}\, .
\end{equation}

\noindent
The constant $\xi$ has to be $+1$ in order to preserve the energy-momentum
tensor. The same conditions can be derived from the requirement that the
matrix $M_{N}{}^{M}$ preserves the off-diagonal metric $(\eta^{MN})=\left(
  \begin{array}{cc}
    0 & \mathbb{I}_{n_{2}\times n_{2}} \\
    \mathbb{I}_{n_{2}\times n_{2}} & 0 \\
  \end{array}
\right)$,
that is

\begin{equation}
M_{M}{}^{P}\eta_{PQ}M_{N}{}^{Q}=\eta_{MN}\, .  
\end{equation}

Thus, the system of 2-form equations of motion and Bianchi identities is
invariant under symmetries that can be embedded into $SO(n_{2},n_{2})$. The
off-diagonal metric $\eta$ can be used to raise and lower $M,N=1,\cdots,2n_{2}
$ indices, in agreement with the definitions (\ref{eq:so2n2n2vectors}) of the
vectors $d^{M}{}_{ij}$ and $d_{M\, ij}$.

Only those transformations of the matrices $b_{\Lambda\Sigma}$ and
$c_{\Lambda\Sigma}$ that can be compensated by a reparametrization of the
scalar manifold leaving invariant the target-space metric $g_{xy}(\phi)$ will
be symmetries of the theory. Furthermore, the reparametrizations of the scalar
manifold must induce linear transformations $M_{i}{}^{j}$ of the 1-forms'
kinetic matrix $a_{ij}(\phi)$ that can be compensated by the inverse linear
transformation acting on the 1-forms.

Defining the  $SO(n_{2},n_{2})$ generators by 

\begin{equation}
M_{M}{}^{N}\sim \delta_{M}{}^{N}  +\alpha^{A}T_{A\, M}{}^{N}\, ,
\end{equation}

\noindent
we find that the above constraint implies 

\begin{equation}
\label{eq:TAantisymm}
T_{A\, (MN)}\equiv T_{A\, (M}{}^{P}\eta_{N)P} =0\, .  
\end{equation}

\noindent
As discussed above, the same transformations must also act linearly on the
1-forms, and, therefore, we can define the generators in the corresponding
representation:

\begin{equation}
M_{i}{}^{j}\sim \delta_{i}{}^{j}  +\alpha^{A}T_{A\, i}{}^{j}\, .
\end{equation}

\noindent
In both representations, the generators $T_{A}$ satisfy the same Lie algebra

\begin{equation}\label{eq:Talgebra}
[T_{A},T_{B}] =-f_{AB}{}^{C}T_{C}\, .  
\end{equation}

\noindent
Since (part of) the symmetry group can act trivially on either vectors or 2-forms we allow some of the generators $T_A$ to be zero. It is for example possible that some symmetry generators act trivially on the 2-forms while they transform some of the scalars and vectors. In this case we have vanishing generators $T_{AM}{}^N$ and non-vanishing $T_{Ai}{}^j$. Still both (formally) satisfy the above algebra.

The $\zeta$-tensor can be redefined in an $SO(n_{2}, n_{2})$-covariant
way:

\begin{equation}
(\zeta^{M}{}_{N})
\equiv
\left(
  \begin{array}{cc}
  0 & \zeta^{\Lambda\Sigma} \\
\zeta_{\Lambda\Sigma} & 0 \\  
  \end{array}
\right)\, ,
\hspace{1cm}
\zeta_{\Lambda\Sigma} = \zeta^{\Lambda\Sigma}\, ,
\end{equation}

\noindent
so the (anti-) self-duality constraint takes the form

\begin{equation}
\zeta^{M}{}_{N}(J^{N}-\zeta^{N}{}_{P}J^{P})=0\, .  
\end{equation}


\subsection{Gaugings and massive deformations}
\label{sec-d6gauging}

In general the above theory will have a group of global symmetries $G$ with
constant parameters $\alpha^{A}$. As discussed in the previous section,
infinitesimally, these global symmetries act on the scalars $\phi^{x}$,
1-forms $A^{i}$ and electric and magnetic 2-forms $B^{M}$ as

\begin{eqnarray}
\delta_{\alpha}\phi^{x} & = & \alpha^{A}k_{A}{}^{x}(\phi)\, ,  \\
& & \nonumber \\
\delta_{\alpha}A^{i} & = & \alpha^{A}T_{A\, j}{}^{i}A^{j}\, , \\
& & \nonumber \\
\delta_{\alpha}B^{M} & = & \alpha^{A}T_{A\, N}{}^{M}B^{N}\, ,
\end{eqnarray}

\noindent
where the matrices $T_{A\, M}{}^{N}$ are generators of
$SO(n_{2},n_{2})$, i.e.~they satisfy Eq.~(\ref{eq:TAantisymm}), and the
$k_{A}{}^{x}(\phi)$ are Killing vectors of the metric $g_{xy}(\phi)$.  Some of
the matrices and Killing vectors may be identically zero. They satisfy the
algebras Eq.~(\ref{eq:Talgebra}) and $[k_A,k_B]=-f_{AB}{}^Ck_C$.

These transformations will be global symmetries of the theory constructed in
the previous section if the following five conditions are met:

\begin{enumerate}
\item The vectors $k_{A}{}^{x}(\phi)$ are Killing vectors of the metric
  $g_{xy}(\phi)$ of the scalar manifold.

\item The kinetic matrices $a_{ij},f_{\Lambda\Sigma}\equiv
  b_{\Lambda\Sigma}+c_{\Lambda\Sigma}$ satisfy the conditions

\begin{eqnarray}
\label{eq:kineticmatrices1}
\pounds_{A}a_{ij} 
& = & 
-2 T_{A\, (i}{}^{k}a_{j)k}\, ,  
\\
& & \nonumber \\
\label{eq:kineticmatrices2}
\pounds_{A}f_{\Lambda\Sigma}
&=& 
-T_{A\, \Lambda\Sigma}
+2T_{A\, (\Lambda}{}^{\Omega}f_{\Sigma) \Omega}
-T_{A}{}^{\Omega\Gamma}f_{\Omega\Lambda}f_{\Gamma\Sigma}\, ,
\end{eqnarray}

\noindent 
where $\pounds_{A}$ denotes the Lie derivative along the vector $k_A$ and the
matrices $T_{A}$ are different components of some of the generators of
$SO(n_{2},n_{2})$ in the fundamental representation 

\begin{equation}
M_{N}{}^{M} \sim \mathbb{I}_{2n_{2}\times 2n_{2}} 
+\alpha^{A}T_{A\, N}{}^{M}
= \mathbb{I}_{2n_{2}\times 2n_{2}}+\alpha^{A}
\left(
  \begin{array}{cc}
   T_{A\, \Sigma}{}^{\Lambda} & T_{A}{}^{\Sigma\Lambda} \\
& \\
   T_{A\, \Sigma\Lambda} & T_{A}{}^{\Sigma}{}_{\Lambda} \\ 
  \end{array}
\right)\, .
\end{equation}

\item The deformation tensor $d_{M\, ij}$ is invariant 

\begin{equation}
\label{eq:YAMij}
\delta_{A}d_{M\, ij} \equiv  Y_{A\, Mij} = -T_{A\, M}{}^{N}d_{N\, ij}
-2T_{A\, (i}{}^{k}d_{M\, j)k}=0\, .    
\end{equation}

\item The scalar potential is invariant

\begin{equation}
\label{eq:potentialagain}
\pounds_{A}V= k_{A}V=0\, .    
\end{equation}

\item The $\zeta$-tensors is invariant

\begin{equation}
\delta_A\zeta^M{}_N = T_{A\, P}{}^{M}\zeta^{P}{}_{N}- T_{A\, N}{}^{P}\zeta^{M}{}_{P}=0\, .
\end{equation}

\end{enumerate}

As we did in the 5-dimensional case, we will relax some of these conditions to
construct a gauged theory. In the next
section when we construct the tensor hierarchy and the action we only require
invariance of $d_{M\,ij}$ under that subgroup of $G$ that is gauged. Taking the limit in
which all deformation tensors but $d_{M\,ij}$ vanish we recover the results of
this section and in particular the action will generically only be invariant
under a subgroup of $G$. The $\zeta$-tensor on the other hand is not a
deformation tensor and we therefore have the condition that it must be an
invariant tensor of the symmetry group.

To gauge the theory we introduce, as in the 5-dimensional case, the embedding
tensor $\vartheta_{i}{}^{A}$, subject to the quadratic constraint
(Eq.~(\ref{eq:QIJA}) with the indices $I,J,K$ replaced by $i,j,k$) which
reflects its gauge-invariance. Following the same steps as in the
5-dimensional case, we introduce the gauge-covariant derivative of the scalars
Eq.~(\ref{eq:covariantderivativescalars}) and, from the Bianchi identity
associated to it, Eq.~(\ref{eq:Bianchiscalars}), we arrive at the definition
of the 2-form field strength $F^{i}$ given in Eq.~(\ref{eq:FI-1}) up to the
undetermined term $\Delta F^{i}$ subject to the condition
Eq.~(\ref{eq:deltaFI-1}). Gauge-covariance of $F^{i}$ implies the gauge
transformation Eq.~(\ref{eq:deltadeltaFI}) for $\Delta F^{i}$, which we
rewrite here for convenience:

\begin{equation}
\delta_{\Lambda}\Delta F^{i}  = -\mathfrak{D}\Delta A^{i}
+2X_{(jk)}{}^{i}[\Lambda^{j}F^{k}+\tfrac{1}{2} A^{j}\wedge \delta_{\Lambda}A^{k}]
\, .
\end{equation}

In this case, in order to satisfy the constraint $\vartheta_{i}{}^{A}\Delta
F^{i}=\vartheta_{i}{}^{A}\Delta A^{i}=0$ it is natural to introduce a matrix
$Z^{iM}$ satisfying

\begin{equation}
\label{eq:QAId6}
Q^{AM}\equiv \vartheta_{i}{}^{A}Z^{iM}=0\, ,  
\end{equation}

\noindent
and define

\begin{equation}
\Delta F^{i} \equiv Z^{iM}B_{M}\, ,
\hspace{1cm}
\Delta A^{i} \equiv -Z^{iM}\Lambda_{M}\, ,
\end{equation}

\noindent
where $\Lambda_{M}$ is the 1-form gauge parameter under which the 2-forms
$B_{M}$ must transform. Then, the gauge transformation of $\Delta F^{i}$
implies

\begin{equation}
Z^{iM}\delta_{\Lambda}B_{M} = Z^{iM} \mathfrak{D}\Lambda_{M} 
+2X_{(jk)}{}^{i}[\Lambda^{j}F^{k}+\tfrac{1}{2} A^{j}\wedge
\delta_{\Lambda}A^{k}]\, .
\end{equation}

\noindent
This solution will only work if $X_{(jk)}{}^{i}\sim Z^{iM} \mathcal{O}_{M\,
  jk}$ for some tensor $\mathcal{O}_{M\, jk}$ symmetric in $jk$. It is natural
to identify this tensor with the tensor $d_{M\, jk}$ that we know can be
introduced in the physical theory so that

\begin{equation}
\delta_{\Lambda}B_{M} = \mathfrak{D}\Lambda_{M} 
+2d_{M\,jk}[\Lambda^{j}F^{k}+\tfrac{1}{2} A^{j}\wedge
\delta_{\Lambda}A^{k}]+\Delta B_M\, ,
\end{equation}

\noindent 
in which $Z^{iM}\Delta B_M=0$. With this choice for we find agreement with
what was found in the previous subsection obtained by setting
$\vartheta_{i}{}^{A}=Z^{iM}=0$.

We impose the constraint

\begin{equation}
\label{eq:QJKId6}
Q_{jk}{}^{i}\equiv X_{(jk)}{}^{i}- Z^{iM} d_{M\, jk}=0\, , 
\end{equation}

\noindent
where we have chosen the normalization of $d_{M\, jk}$ to recover the
expression we got in the previous section. We thus find

\begin{eqnarray}
F^{i} 
& = & 
dA^{i}+\tfrac{1}{2}X_{jk}{}^{i}A^{jk}  +Z^{iM}B_{M}\, ,\\
& & \nonumber \\
\delta_{\Lambda}A^{i} 
& = &  
-\mathfrak{D}\Lambda^{i} -Z^{iM}\Lambda_{M}\, ,\\
& & \nonumber \\
\delta_{\Lambda}B_{M} 
& = & 
 \mathfrak{D}\Lambda_{M} 
+2 d_{M\, kl}(\Lambda^{k}F^{l} +\tfrac{1}{2}A^{k}\wedge
\delta_{\Lambda}A^{l} ) +\Delta B_{M}\, ,
\hspace{.4cm}
 Z^{iM}\Delta B_{M}=0\, ,
\end{eqnarray}

\noindent
where the possible additional term $\Delta B_{M}$ will be determined by the
requirement of gauge-covariance of the 3-form field strength $H_{M}$.

We must require the tensors $Z^{iM}$ and $d_{M\, ij}$ to be gauge-invariant, which
leads to the constraints

\begin{eqnarray}
Q_{i}{}^{jM} 
& \equiv & 
-\delta_{i}Z^{jM}
=
-X_{ik}{}^{j}Z^{kM} -X_{iN}{}^{M}Z^{jN} =0\, ,\label{eq:gaugeinvZ}\\
& & \nonumber \\
Q_{i\, M\, jk} 
& \equiv & 
-\delta_{i} d_{M\, jk}
=
X_{i\, M}{}^{N}d_{N\, jk}+2X_{i\, (j|}{}^{l}d_{M\, |k)l}
=0\, .  
\end{eqnarray}

\noindent
This last constraint is clearly weaker than the global invariance constraint
$Y_{A\, Mij}=0$ in Eq.~(\ref{eq:YAMij}).


\subsubsection{The 3-form field strengths $H_{M}$}
\label{sec-d6HM}

The covariant derivative of the 2-form field strengths $F^{i}$, after use of
the generalized Jacobi identities\footnote{In the 6-dimensional theory the generalized Jacobi identity reads $X_{[jk}{}^mX_{l]m}{}^i=\tfrac{2}{3}Z^i{}_NX_{[jk}{}^md^N{}_{l]m}$.} is

\begin{equation}
\mathfrak{D}F^{i} 
=  
Z^{iM}\{\mathfrak{D}B_{M}+d_{M\, jk}[A^{j}\wedge dA^{k}
+\tfrac{1}{3} X_{lm}{}^{k}A^{jlm}]\}\, ,
\end{equation}

\noindent
which leads us to define the 3-form field strength

\begin{eqnarray}
\label{eq:BianchiFId6}
\mathfrak{D}F^{i} 
& = & 
Z^{iM} H_{M}\, ,
\\
& & \nonumber \\
H_{M} 
& \equiv & 
\mathfrak{D}B_{M}+ d_{M\, jk}[A^{j}\wedge dA^{k}
+\tfrac{1}{3}X_{lm}{}^{k}A^{jlm}]  +\Delta H_{M}\, ,
\\
& & \nonumber \\
Z^{iM}\Delta H_{M} & = & 0\, ,
\end{eqnarray}

\noindent
where $\Delta H_{M}$ will be determined, together with $\Delta B_{M}$ by using
gauge-covariance of $H_{M}$, which is guaranteed by the formalism. To proceed
with constructing the hierarchy we do not need the explicit form of the gauge
transformations $\Delta B_{M}$. Just as in the 5-dimensional case we can
continue with constructing gauge-covariant field strengths by computing the
Bianchi identities. The form of $\Delta H_{M}$ will be a contraction of some
invariant tensor(s), that are annihilated by $Z^{iM}$, with some 3-forms. We
will determine $\Delta H_M$ simultaneously with the 4-form field strengths
$G_i$.


\subsubsection{The 4-form field strengths $G_{i}$}
\label{sec-d6Gi}

The Bianchi identity of $H_{M}$ takes the form

\begin{equation}
  \begin{array}{rcl}
\mathfrak{D} H_{M} & = & 
d_{M\, ij} F^{ij} 
+\mathfrak{D}\Delta H_{M}
\\
& & \\
& & 
+Z_{M\, i}{}^{N}
\left\{ 
(F^{i} -\tfrac{1}{2}Z^{iP}B_{P})\wedge B_{N}
+\tfrac{1}{3}d_{N\, jk}A^{ij}\wedge dA^{k}
+\tfrac{1}{12} X_{jk}{}^{n}d_{N\, ln}A^{ijkl}
\right\}  
\, ,\\
\end{array}
\end{equation}

\noindent
where we have defined the tensor

\begin{equation}
Z_{M\, i}{}^{N} \equiv 
-X_{i\, M}{}^{N} -2d_{M\, ij}Z^{jN}\, ,  
\end{equation}

\noindent
which is annihilated by $Z^{jM}$, i.e.~$Z^{jM}Z_{M\, i}{}^{N}=0$ by virtue of Eqs.~(\ref{eq:QAId6}), (\ref{eq:QJKId6}) and (\ref{eq:gaugeinvZ}).

The simplest Ansatz we can make is to assume that $\Delta H_{M} = Z_{M\,
  i}{}^{N}C_{N}{}^{i}$ for some 3-forms $C_{N}{}^{i}$. However, in $d=6$
dimensions the 3-forms of a physical theory are dual to the 1-forms, and,
therefore, as we have shown in the case that $\vartheta_i{}^A=Z^{iM}=0$, we
can only have 3-forms $C_{i}$. This means that we must define a new\footnote{
\label{foot:esa}
  In principle $Z_{M}{}^{i}$ and $Z^{i}{}_{M}$ are unrelated, but we are going to see that we can
  relate these two tensors, though. This is not just an economical possibility,
  but reflects the fact that if a $p$-form has a st\"uckelberg coupling to a
  $(p+1)$-form, then their duals, which will be, respectively, $(\tilde{p}+1)$-
  and $\tilde{p}$-forms (with $\tilde{p}=d-p-2$), will also have St\"uckelberg
  couplings with the same parameters and reversed roles: the $\tilde{p}$-form,
  dual of the $(p+1)$-form, will be the St\"uckelberg field of the
  $(\tilde{p}+1)$-form, dual of the $p$-form.} invariant tensor $Z_{M}{}^{i}$
such that

\begin{equation}
\Delta H_{M}  =Z_{M}{}^{i} C_{i}\, ,
\hspace{1cm}
Z^{jM}Z_{M}{}^{i}=0\, .
\end{equation}

In order to make contact with the field strength $G_{i}$ in
Eq.~(\ref{eq:redefinedGi}) of the theory obtained for
$\vartheta_i{}^A=Z^{iM}=0$ we must require

\begin{equation}\label{eq:3formcondition}
Z_{M\, i}{}^{N} = 2Z_{M}{}^{j} d^{N}{}_{ji}\, ,  
\end{equation}

\noindent
so that the Bianchi identity will take the form

\begin{equation}
  \begin{array}{rcl}
\mathfrak{D} H_{M} 
& = & 
d_{M\, ij} F^{i}\wedge F^{j} 
+Z_{M}{}^{i} G_{i}\, ,
\\
& & \\
G_{i}
& = & 
\mathfrak{D}C_{i}
+2d^{N}{}_{ip}
\left[
(F^{p} -\tfrac{1}{2}Z^{pM}B_{M})\wedge B_{N}
+\tfrac{1}{3}d_{N\, jk} A^{pj}\wedge dA^{k}
+\tfrac{1}{12} X_{jk}{}^{n}d_{N\, ln}A^{pjkl}
\right]
\\
& & \\
& & 
+\Delta G_{i}
\, ,
\\
& & \\
Z_{M}{}^{i}\Delta G_{i} 
& = & 
0\, .\\
\end{array}
\end{equation}

\noindent
The requirement~(\ref{eq:3formcondition}) leads to 

\begin{equation}
X_{i\, MN} = -2(d_{M\, ij}Z^{j}{}_{N}+ d_{N\, ij}Z_{M}{}^{j})\, .  
\end{equation}

\noindent
The antisymmetry of $X_{i\, MN}$ suggests\footnote{See footnote~\ref{foot:esa}.} to take

\begin{equation}
Z^{Mi}=-Z^{iM}\, . 
\end{equation}

Summarizing we have thus two new constraints: 

\begin{eqnarray}
Q_{i\, MN}
& \equiv & 
X_{i\, MN} - 4Z^{j}{}_{[M}d_{N] ij}=0\, ,
\\
& & \nonumber \\
Q^{ij} 
& \equiv & 
Z^{iM}Z^{j}{}_{M} =0\, ,  
\end{eqnarray}

\noindent
from which it follows that the tensor

\begin{equation}
C_{MNP}\equiv d_{M\, ij} Z^{i}{}_{N}Z^{j}{}_{P}\, , 
\end{equation}

\noindent
is totally symmetric. 

The constraint $Q^{ij}=0$ is similar to the constraint
$\vartheta_{M}{}^{A}\vartheta^{MB}=0$ in 4 dimensions
\cite{Bergshoeff:2009ph}.

We will show the validity of this construction by proving the consistency
of the resulting tensor hierarchy.


\subsubsection{The 5-form field strengths $K_{A}$}
\label{sec-d6KA}

If we take the covariant derivative of the Bianchi identity of $H_{M}$
we find 

\begin{equation}
Z_{M}{}^{i}[\mathfrak{D}G_{i} -2 d^{N}{}_{ij}F^{j}\wedge H_{N}] =0\, ,  
\end{equation}

\noindent
from which it follows that the Bianchi identity of $G_{i}$ must have the form

\begin{equation}
\mathfrak{D}G_{i} =2 d^{N}{}_{ij}F^{j}\wedge H_{N}  +\Delta \mathfrak{D} G_{i}\, ,
\hspace{1cm}
Z_{M}{}^{i}\Delta \mathfrak{D}G_{i}=0\, .
\end{equation}

A direct calculation using the above expression for $G_{i}$ gives the result

\begin{equation}
  \begin{array}{rcl}
\mathfrak{D}G_{i} 
& = & 
2 d^{M}{}_{ij}F^{j}\wedge H_{M}  + \mathfrak{D} \Delta G_{i}
+\vartheta_{i}{}^{A}
\left\{
T_{A}{}^{MN}(H_{M}-\tfrac{1}{2}\mathfrak{D}B_{M})\wedge B_{N}
\right.
\\
& & \\
& & 
+T_{A\, k}{}^{p}
\left[
(F^{k}-Z^{kM}B_{M})\wedge C_{p}
-\tfrac{1}{6}
d^{M}{}_{jp}d_{M\, lm} A^{jkl}\wedge dA^{m}
\right.
\\
& & \\
& & 
\left.
\left.
+\tfrac{1}{30}X_{lm}{}^{q}d^{M}{}_{jq}d_{M\, pn} A^{jklmn}
\right]
\right\}\, ,  
\end{array}
\end{equation}

\noindent
up to terms proportional to the constraint Eq.~(\ref{eq:ddsonn}) which, so far
we had not needed. The reason why we need to use it here is that the term
$d_{M\, i(j} d^{M}{}_{kl)}$ is not annihilated by $Z^{iN}$ and we cannot argue
that it is proportional to $\vartheta_{i}{}^{A}$ times some new tensor. the
only consistent way forward is to use Eq.~(\ref{eq:ddsonn}).

Since $Z^{iM}\vartheta_{i}{}^{A}=0$, we can set $\Delta
G_{i}=\vartheta_{i}{}^{A}D_{A}$ for some 4-forms $D_{A}$ and write the Bianchi
identity for the 4-form field strength $G_{i}$ in the form

\begin{eqnarray}
\mathfrak{D}G_{i} 
& = & 
2 d^{M}{}_{ij}F^{j}\wedge H_{M}  +\vartheta_{i}{}^{A}K_{A}\, ,
\\
& & \nonumber \\
K_{A} 
& = & 
\mathfrak{D}D_{A} 
+T_{A}{}^{MN}(H_{M}-\tfrac{1}{2}\mathfrak{D}B_{M})\wedge B_{N}
\nonumber \\
& & \nonumber \\
& & 
+T_{A\, k}{}^{p}
\left[
(F^{k}-Z^{kM}B_{M})\wedge C_{p}
-\tfrac{1}{6}
d^{M}{}_{jp}d_{M\, lm} A^{jkl}\wedge dA^{m}
+\tfrac{1}{30}X_{lm}{}^{q}d^{M}{}_{jq}d_{M\, pn} A^{jklmn}
\right]
\nonumber \\
& & \nonumber \\
& & 
+\Delta K_{A}\, ,\\
& & \nonumber \\
\vartheta_{i}{}^{A}\Delta K_{A} 
& = & 0\, .
\end{eqnarray}


\subsubsection{The 6-form field strengths $L$}
\label{sec-d6L}

The covariant derivative of the Bianchi identity of $G_{i}$ implies that the
Bianchi identity for the 5-form field strengths must be of the form 

\begin{equation}
\mathfrak{D}K_{A} = 
T_{A\, j}{}^{k} F^{j}\wedge G_{k}   
-\tfrac{1}{2} T_{A}{}^{MN}H_{M}\wedge H_{N} +\Delta \mathfrak{D}K_{A}\, ,
\hspace{1cm}
\vartheta_{i}{}^{A}\Delta \mathfrak{D}K_{A}=0\, .
\end{equation}

It is useful to have some idea of what we can expect concerning
$\mathfrak{D}K_{A}$ according to the general formalism that we have introduced
before.

As we have seen, 6-dimensional gauge theories are determined by three
different deformation tensors $\vartheta_{i}{}^{A},Z^{iM},d_{M\, ij}$
satisfying the 5 constraints $Q=0$:

\begin{eqnarray}
Q^{AM} 
& \equiv  & 
\vartheta_{i}{}^{A}Z^{iM}\, ,
\\
& & \nonumber \\
Q^{ij} 
& \equiv & 
Z^{iM}Z^{j}{}_{M} \, ,
\\
& & \nonumber \\
Q_{jk}{}^{i}
& \equiv  & 
X_{(jk)}{}^{i} -Z^{iM}d_{M\, jk}\, ,
\\
& & \nonumber \\
Q_{i\, MN}
& \equiv & 
X_{i\, MN} - 4Z^j{}_{[M}d_{N] ij}\, ,
\\
& & \nonumber \\
Q_{ijk\, l} 
& \equiv  & 
d_{M\, (ij}d^{M}{}_{k)l}\, ,
\end{eqnarray}

\noindent
plus the three constraints associated to the gauge-invariance of the deformation
tensors:

\begin{eqnarray}
Q_{ji}{}^{A}
& \equiv &
-\delta_{j} \vartheta_{i}{}^{A}
=
-\vartheta_{j}{}^{B}Y_{B\, i}{}^{A} 
= 
-\vartheta_{j}{}^{B}
(f_{BC}{}^{A}\vartheta_{i}{}^{C}-T_{B\, i}{}^{k}\vartheta_{k}{}^{A})\, ,
\\
& & \nonumber \\
Q_{j}{}^{iM}
& \equiv & 
-\delta_{j} Z^{iM}
=
-\vartheta_{j}{}^{A} Y_{A}{}^{iM}
=
-\vartheta_{j}{}^{A}
(T_{A\, k}{}^{i} Z^{kM} +T_{A\, N}{}^{M}Z^{iN})\, ,
\\
& & \nonumber \\
Q_{k\, M\, ij}
& \equiv & 
-\delta_{k}d_{M\, ij}
=
-\vartheta_{k}{}^{A}
Y_{A\, M\, ij}
=
\vartheta_{k}{}^{A}
(2T_{A\, (i|}{}^{l}d_{M\, |j)l} +T_{A\, M}{}^{N}d_{N\, ij})\, .
\end{eqnarray}

We thus expect three 5-forms $E^{i}{}_{A},E_{iM},E^{M\, ij}$ dual to the
deformation tensors that will appear in the field strength $K_{A}$ through the
term

\begin{equation}
\Delta K_{A} =   
Y_{A\, i}{}^{B}E^{i}{}_{B}
+Y_{A}{}^{iM}E_{iM}
+Y_{A\, M\, ij}E^{M\, ij}\, .
\end{equation}

The result of a direct calculation is

\begin{equation}
  \begin{array}{rcl}
\mathfrak{D}K_{A} 
& = &  
T_{A\, j}{}^{k} F^{j}\wedge G_{k}   
-\tfrac{1}{2} T_{A}{}^{MN}H_{MN} 
\\
& & \\
& & 
+Y_{A\, i}{}^{B}
\{-F^{i}\wedge D_{B}
+\tfrac{1}{30} T_{B\, k}{}^{n}d^{N}{}_{jm}d_{N\, ln}A^{ijkl}\wedge dA^{m} 
+\tfrac{1}{80}  T_{B\, k}{}^{p} X_{lm}{}^{q} 
d^{N}{}_{jq}d_{N\, pn}A^{ijklmn}\}
\\
& & \\
& & 
+Y_{A}{}^{iM}
\{(H_{M}-\mathfrak{D}B_{M})\wedge C_{i} 
-B_{M}\wedge (G_{i}-\vartheta_{i}{}^{B}D_{B}) 
-\tfrac{1}{2}Z^{j}{}_{M}C_{ij}
\\
& & \\
& & 
+d^N{}_{ij}F^{j}\wedge B_{MN} +\tfrac{1}{3}d^N{}_{ij}Z^{jP}B_{MNP} \}\\
& & \\
& & 
+Y_{A}{}^{M}{}_{ij}
\{-F^{ij} \wedge B_{M} +Z^{iN}F^{j} \wedge B_{MN}
-\tfrac{1}{3}Z^{iN}Z^{jP}B_{MNP}
\\
& & \\
& & 
-\tfrac{1}{2} d_{M\, kl}A^{ik}\wedge d A^{jl} 
-\tfrac{2}{15} X_{kl}{}^{n}d_{M\, nm} A^{iklm}\wedge dA^{j}
-\tfrac{1}{5}X_{kl}{}^{j} d_{M\, nm} A^{ikln}\wedge dA^{m}
\\
& & \\
& & 
-\tfrac{1}{18}X_{kl}{}^{j} X_{np}{}^{q} d^M{}_{mq}A^{iklmnp}\}
\\
& & \\
& & 
+\mathfrak{D}\Delta K_{A}\, .
\\
\end{array}
\end{equation}

\noindent
If we take $\Delta\mathfrak{D}K_A$ to be 

\begin{equation}
 \Delta\mathfrak{D}K_A=Y_{A\, i}{}^{B}L^{i}{}_{B}
+Y_{A}{}^{iM}L_{iM}
+Y_{A\, M\, ij}L^{M\, ij}\, ,
\end{equation}

\noindent
where $L^{i}{}_{B}$, $L_{iM}$, $L^{M\, ij}$ are the gauge-covariant field
strengths of the 5-forms $E^{i}{}_{B}$, $E_{iM}$ and $E^{M\, ij}$,
respectively, then we obtain the Bianchi identity for $K_A$ given in
Eq.~(\ref{eq:BianchiKA}) with the 6-form field strengths $L^{i}{}_{B}$,
$L_{iM}$, $L^{M\, ij}$ given in Eqs.~(\ref{eq:LiB}), (\ref{eq:LiM}) and
(\ref{eq:LMij}).

In Eqs.~(\ref{eq:LiB}), (\ref{eq:LiM}) and (\ref{eq:LMij}) we have not
specified in detail the St\"uckelberg couplings to the 6-forms that we denoted
by $F_\flat$. There are in total eight top-forms in 6-dimensions corresponding
to the eight constraints. These eight top-forms are determined up to massive
gauge transformations of the form $\delta F_\flat=\Sigma_\flat$ such that
$W_\sharp{}^\flat\Sigma_\flat=0$. This is because all the top-forms only come
contracted with $W_\sharp{}^\flat$. In particular theories it can happen that
these massive gauge transformations enable one to complete gauge away certain
top-forms entirely. The massless gauge transformations of the top-forms
contain the 5-form gauge transformation parameter $\Lambda_\flat$,
i.e. $W_\sharp{}^\flat\delta
F_\flat=W_\sharp{}^\flat\mathfrak{D}\Lambda_\flat$. This parameter also shows
up in the gauge transformation of the 5-form potentials $E_\sharp$ as $\delta
E_\sharp=-W_\sharp{}^\flat\Lambda_\flat$. Depending on the details of the
theory these massive gauge transformation may allow one to entirely gauge away
certain 5-forms.


\subsubsection{Gauge-invariant action for the 1-, 2- and 3-forms}
\label{sec-d6action}

Our starting point to construct a 6-dimensional gauge-invariant action
is\footnote{We do not consider the Einstein-Hilbert term as it plays no role in the
  discussion.}

\begin{equation}
\begin{array}{rcl}
S_{1} 
& \equiv & 
{\displaystyle\int} 
\left\{
\tfrac{1}{2}g_{xy}(\phi) \mathfrak{D}\phi^{x}\wedge \star \mathfrak{D}\phi^{y}
-\tfrac{1}{2} a_{ij}(\phi)F^{i}\wedge \star F^{j}
\right.
\\
& & \\
& & 
\left.
+ \tfrac{1}{2} b_{\Lambda\Sigma}(\phi) H^{\Lambda}\wedge \star H^{\Sigma}
+ \tfrac{1}{2} c_{\Lambda\Sigma}(\phi) H^{\Lambda}\wedge H^{\Sigma}
+\star V(\phi)
\right\}\, ,    
\end{array}
\end{equation}

\noindent
where the covariant derivative and field strengths are those of the tensor
hierarchy. This means, in particular, that 

\begin{equation}
\mathfrak{D}B^{\Sigma}=  dB^{\Sigma} +X_{i\, M}{}^{\Sigma}A^i\wedge B^{M}\, ,
\end{equation}

\noindent
so the magnetic 2-forms $B_{\Sigma}$ occur in this action.

As a general rule, the gauge-invariant action will only differ from this one
by topological Chern--Simons-like terms. Furthermore, the equations of motion
will just be gauge-covariant generalizations of the ungauged ones, up to
duality transformations. More precisely, as a general rule, the equations of motion
of the magnetic higher-rank form fields (here the magnetic 2-forms
$B_{\Sigma}$ and the 3-forms $C_{i}$) will just be duality relations, and the
equations of motion of the (electric) lower-rank potentials (here the 1-forms
$A^{i}$ and the electric 2-forms $B^{\Sigma}$) will be completely equivalent
to the hierarchy's Bianchi identities after use of the duality relations.

Let us first consider all those which contain the 3-forms $C_{i}$. Taking into
account that we expect the equation of motion of $C_{i}$ to be a duality
relation for the 3-form field strengths, a reasonable Ansatz for the terms
that involve 3-forms is

\begin{equation}
S_{2} \equiv   {\displaystyle\int} 
Z^{i\Sigma}C_{i}\wedge (H_{\Sigma} +\tfrac{1}{2}Z^{j}{}_{\Sigma}C_{j})\, ,
\end{equation}

\noindent
since, if we only vary w.r.t.~the 3-forms, we get 

\begin{equation}
\label{eq:3formvariation}
\delta (S_{1}+S_{2}) =  -Z^{iM}\delta C_{i}\wedge
[J_{M}-H_{M}]\, , 
\end{equation}
 
\noindent
where $J_{\Lambda}$  is given in Eq.~(\ref{eq:JL}) (but with the field strengths $H^\Lambda$ replaced by those of the hierarchy) and the upper component of the doublet $J^{M}$ is defined to be
$J^{\Sigma}\equiv H^{\Sigma}$.

Let us now consider the topological terms containing magnetic 2-forms
$B_{\Lambda}$. We expect the equations of motion of the $B_{\Lambda}$ to give
the duality relation between 2- and 4-form field strengths (up to, possibly,
other duality relations). If we only vary $B_{\Sigma}$ in $S_{1}+S_{2}$ we
find the result

\begin{equation}
\delta (S_{1}+S_{2}) 
=
\delta B_{\Sigma} \wedge 
\left\{
-Z^{i\Sigma}[a_{ij}\star F^{j} -\mathfrak{D}C_{i}]
+X_{i}{}^{\Sigma\Omega} A^{i} \wedge 
[J_{\Omega} +Z^{j}{}_{\Omega}C_{j}] 
\right\}\, ,
\end{equation}

\noindent
whose two terms have the form of incomplete duality relations, in agreement
with our prejudice. If we require that the next term we add to the action,
$S_{3}$, gives, upon variation of $B_{\Sigma}$ only, the complete duality
relations

\begin{equation}
\label{eq:magnetic2formvariation}
\delta (S_{1}+S_{2}+S_{3}) 
=
-\delta B_{\Sigma} \wedge 
\left\{
Z^{i\Sigma}[a_{ij}\star F^{j} -G_{i}]
+\mathfrak{D}(J^{\Sigma}-H^\Sigma)] 
\right\}\, ,
\end{equation}

\noindent
we find that 

\begin{equation}
\begin{array}{rcl}
  S_{3}
  & \equiv & 
  {\displaystyle\int} 
  \biggl \{
  B_{\Sigma} \wedge 
  \left\{
    Z^{i\Sigma}
    [2d_{\Omega\, ij} f^{j}\wedge B^{\Omega} 
    +g_{i}
    +\tfrac{1}{2} d^{\Omega}{}_{ij}X_{kl}{}^{j} A^{kl}\wedge B_{\Omega}
    ]
  \right.
  \\
  & & \\
  & & 
  \left.
    +2d^{\Sigma}{}_{ij} Z^{j\Omega} A^{i}\wedge d B_{\Omega}
    +X_{i}{}^{\Sigma\Omega} A^{i}\wedge 
    [-h_{\Omega} +X_{j\, \Omega\Gamma} A^{j}\wedge B^{\Gamma} 
    +\tfrac{1}{2} X_{j\, \Omega}{}^{\Gamma} A^{j}\wedge B_{\Gamma} ]
  \right\}
  \\
  & & \\
  & & 
  +\tfrac{1}{3}d^{M}{}_{ij}Z^{iN}Z^{jP} B_{MNP}
  -\tfrac{1}{3}d_{\Lambda\, ij}Z^{i}{}_{\Sigma}Z^{j}{}_{\Omega}
  B^{\Lambda\Sigma\Omega}
\biggr \}\, ,
\end{array}
\end{equation}

\noindent
where $f^j$, $h_{\Omega}$ and $g_{i}$ are, respectively, the part of the field
strengths $F^j$, $H_{\Omega}$ and $G_{i}$ that only depend on the 1-forms
$A^{i}$, i.e.

\begin{eqnarray}
f^j & \equiv & dA^j+\tfrac{1}{2}X_{kl}{}^jA^{kl}\, ,
\\
& & \nonumber \\
h_{M} 
& \equiv & 
d_{M\,jm}A^{j}\wedge dA^{m}
+\tfrac{1}{3}d_{M\, jm}X_{kl}{}^{m}A^{jkl}\, ,
\\
& & \nonumber \\
g_{i}
& \equiv &
\tfrac{2}{3}d^{M}{}_{ij}d_{M\, kl} A^{jk}\wedge dA^{l}
+\tfrac{1}{6} d^{M}{}_{ij}d _{M\, kl}X_{mn}{}^{l}A^{jkmn}
\, .
\end{eqnarray}

Observe that $S_{3}$ does not contain any 3-forms and, therefore, the
variation of the action w.r.t.~the 3-forms, Eq.~(\ref{eq:3formvariation}), does
not change when we add $S_{3}$.

We next consider the variations w.r.t.~the electric 2-forms
$B^{\Sigma}$. These should give the equations of motion of the electric
2-forms  up to duality relations. Adding

\begin{equation}
\begin{array}{rcl}
S_{4}
& \equiv & 
{\displaystyle \int}
\biggl \{
d_{\Sigma\, ij}B^{\Sigma}\wedge f^{ij}
+\tfrac{1}{3}d_{\Lambda\, ij}Z^{i}{}_{\Sigma}Z^{j}{}_{\Omega} 
B^{\Lambda\Sigma\Omega}
\\
& & \\
& & 
+X_{i\, \Sigma\Omega} A^{i} \wedge h^{\Omega} \wedge B^{\Sigma}
+2d_{\Sigma\, ij} Z^{i}{}_{\Omega} A^{j}\wedge dB^{\Sigma}\wedge B^{\Omega}
\\
& & \\
& & 
+\tfrac{1}{2}
(d_{\Sigma\, ij}Z^{i}{}_{\Omega} X_{kl}{}^{j} -X_{k\,
  \Sigma\Gamma}X_{l}{}^{\Gamma}{}_{\Omega})
A^{kl}\wedge B^{\Sigma\Omega}
\biggr \}\, ,   
\end{array}
\end{equation}

\noindent
we find that varying only w.r.t.~$B^{\Sigma}$ gives

\begin{equation}
\delta (S_{1}+S_{2}+S_{3}+S_{4}) 
=
-\delta B^{\Sigma} \wedge   
\{Z^{i}{}_{\Sigma}[a_{ij}\star F^{j} -G_{i}]
+\mathfrak{D}(J_{\Sigma}-H_\Sigma)\}\, ,
\end{equation}

\noindent
which, upon duality relations gives the hierarchy's Bianchi identity of the
magnetic 3-form field strengths $H_{\Sigma}$.  $S_{4}$ does not contain any
3-forms or magnetic 2-forms and, therefore, adding $S_{4}$ does not change
neither Eq.~(\ref{eq:3formvariation}) nor
Eq.~(\ref{eq:magnetic2formvariation}).

Finally, let us consider the variation of $S_{1}$ w.r.t.~the 1-forms
$A^{i}$ only. We can write the result in the form

\begin{equation}
\delta S_{1} =
\delta A^{i} \wedge \biggl \{
-\star{\displaystyle\frac{\delta S}{\delta A^{i}}} 
+s_{i} \biggr \}\, ,
\end{equation}

\noindent
where we have \textit{defined} 

\begin{equation}
  \begin{array}{rcl}
\star{\displaystyle\frac{\delta S}{\delta A^{i}}} 
& \equiv &
\mathfrak{D}(a_{ij}\star F^{j}) -2 d^{M}{}_{ij}F^{j}\wedge J_{M}
-\vartheta_{i}{}^{A}\star j_{A}
\\
& & \\
& & 
+d^{M}{}_{ij}A^{j}\wedge \left[Z^{k}{}_M(a_{kl}\star F^{l} -G_{k})
+\mathfrak{D}(J_{M} -H_M)\right]
\\
& & \\
& & 
+[2 d_{N\, il} B^{N} +\tfrac{2}{3} d^{N}{}_{lj} d_{N\,
  ki}A^{jk}] \wedge Z^{lM}[J_{M}-H_{M}]\, ,
\end{array}
\end{equation}

\noindent
and

\begin{equation}
  \begin{array}{rcl}
s_{i} & \equiv & 
-d_{\Sigma\, ij}Z^{k\Sigma}A^{j}\wedge G_{k}
-(2d_{\Sigma\, ij} F^{j} +d^{\Omega}{}_{ij} X_{k\, \Omega\Sigma}A^{jk})\wedge H^{\Sigma}
\\
& & \\
& & 
+[X_{i\, M}{}^{\Sigma}B^{M} +d^{\Sigma}{}_{l[i}X_{jk]}{}^{l} A^{jk} 
-d^{\Omega}{}_{ij} X_{k\, \Omega}{}^{\Sigma}A^{jk}
-2 d^{\Sigma}{}_{ij}(F^{j}-dA^{j})]\wedge H_{\Sigma}
\\
& & \\
& & 
-d^{\Sigma}{}_{ij}d_{\Sigma\, kl}A^{j}\wedge F^{kl}
\, .
\end{array}
\end{equation}

\noindent
While this definition is mainly based on intuition, we can check that the
variations of the pieces $S_{2},S_{3}$ and $S_{4}$ w.r.t.~$A^{i}$ only
contribute to $s_{i}$: the variation of $S_{2}$ w.r.t.~$A^{i}$ cancels all the
terms in $s_{i}$ containing the 3-forms $C_{i}$; the variation of $S_{3}$
w.r.t.~$A^{i}$ cancels all the terms in $s_{i}$ containing the magnetic
2-forms $B_{\Sigma}$ and the variation of $S_{4}$ w.r.t.~$A^{i}$ cancels all
the terms in $s_{i}$ containing the electric 2-forms $B^{\Sigma}$, leaving
unchanged what we have defined as ${\displaystyle\frac{\delta S}{\delta
    A^{i}}}$. Thus, we only need to see if there exists an $S_{5}$ whose
variation w.r.t.~$A^{i}$ cancels the terms in $s_{i}$ that only depend on the
1-forms $A^{i}$. In other words: we have to determine the integrability of the
terms in $\delta A^{i}\wedge s_{i}$ that only depend on 1-forms. This highly
non-trivial requirement is satisfied and $S_{5}$ is given by

\begin{equation}
  \begin{array}{rcl}
S_{5}
& = & 
\tfrac{1}{4} [d_{\Sigma\, ik}d^{\Sigma}{}_{jl} -d^{\Sigma}{}_{ik}d_{\Sigma\,
  jl}] A^{ij}\wedge dA^{kl}     
\\
& & \\
& & 
+X_{ij}{}^{p}[\tfrac{2}{15}d_{\Sigma\, km}d^{\Sigma}{}_{lp}
-\tfrac{1}{5}d^{\Sigma}{}_{km}d_{\Sigma\, lp}] A^{ijkl}\wedge dA^{m}
\\
& & \\
& & 
+\tfrac{1}{9}[ d_{\Sigma\, ip}d^{\Sigma}{}_{jq}
+\tfrac{1}{2}d^{\Sigma}{}_{ip}d_{\Sigma\, jq}]
X_{kl}{}^{p}X_{mn}{}^{q}A^{ijklmn}\, .
  \end{array}
\end{equation}

\noindent
It is evident that this additional term does not modify the variations of the
total action\footnote{A similar action for the case of the maximal 6-dimensional supergravity theory was constructed in \cite{Bergshoeff:2007ef}.} 

\begin{equation}
S \equiv S_{1}+\cdots +S_{5} 
\end{equation}

\noindent
w.r.t.~the 3- and 2-forms. 

We, thus arrive at the following result:

\begin{equation}
  \begin{array}{rcl}
\delta S 
& = & 
{\displaystyle \int} 
\biggl \{
-\delta \phi^{x} \star {\displaystyle \frac{\delta S}{\delta \phi^{x}}}
-\delta A^{i}\wedge \star {\displaystyle \frac{\widetilde{\delta S}}{\delta A^{i}}}
-(\delta B^{M}-d^{M}{}_{ij}A^{i}\wedge \delta A^{j})\wedge 
\star {\displaystyle \frac{\delta S}{\delta B^{M}} }
\\
& & \\
& & 
-[
\delta C_{i} + 2d_{M\, ij} B^{M}\wedge \delta A^{j}
+\tfrac{2}{3} d^{M}{}_{ij}d_{M\, kl}A^{jk}\wedge \delta A^{l}
]\wedge {\displaystyle \frac{\delta S}{\delta C_{i}}}
\biggr \}  \, ,
\end{array}
\end{equation}

\noindent
where 

\begin{eqnarray}
\star \frac{\delta S}{\delta \phi^{x}}
& = & 
g_{xy}\mathfrak{D}\star\mathfrak{D}\phi^{y}
+\tfrac{1}{2} \partial_{x}a_{ij}F^{i}\wedge \star F^{j}
-\tfrac{1}{2} H^{M}\wedge \partial_{x}J_{M}
-\star \partial_{x}V\, ,
\\
& & \nonumber \\
\frac{\delta S}{\delta C_{i}}
& = & 
Z^{iM}(J_{M}-H_{M})\, ,
\\
& & \nonumber \\
\star \frac{\delta S}{\delta B^{M}}  
& = & 
Z^{i}{}_{M} (a_{ij}\star F^{j}-G_{i})+\mathfrak{D}(J_M-H_M)\, ,
\\
& & \nonumber \\
 \star \frac{\widetilde{\delta S}}{\delta A^{i}} 
& = & 
\mathfrak{D}(a_{ij}\star F^{j}) -2 d^{M}{}_{ij}F^{j}\wedge J_{M}
-\vartheta_{i}{}^{A}\star j_{A}\, .
\end{eqnarray}

We can now relate the equations of motion derived from this action and the
tensor hierarchy's Bianchi identities via the duality relations

\begin{eqnarray}
a_{ij}\star F^{j} 
& = & 
G_{i}\, ,
\\
& & \nonumber \\
J_{M}
& = & 
H_{M}\, ,
\\
& & \nonumber \\
\star j_{A}
& = & 
K_{A}\, ,
\\
& & \nonumber \\
 \star \frac{\partial V}{\partial c^{\sharp}}
& = & 
L_{\sharp}\, .
\end{eqnarray}

With these duality relations, the 3-form and magnetic 2-form equations of
motion are automatically solved. The electric 2-form equations of motion
become the hierarchy Bianchi identity of the magnetic 2-forms. The 1-form
equations of motion become the hierarchy's Bianchi identity of the 4-form
field strengths $G_{i}$.  The projected scalar equations of motion
$k_{A}{}^{x} \star{\displaystyle \frac{\delta S}{\delta \phi^{x}}}$ become the
hierarchy's Bianchi identity of the 5-form field strengths $K_{A}$ if we use
that $k_Aa_{ij}=-2T_{A\,(i}{}^ka_{j)k}$ as well as $H^M\wedge k_AJ_M=-T_{A\,M}{}^NJ^M\wedge J_N$,
the Killing property of the $k_{A}{}^{x}$ and the fact that

\begin{equation}
k_{A}V 
= 
\sum_{\sharp}Y_{A}{}^{\sharp}\frac{\partial V}{\partial c^{\sharp}}\, .
\end{equation}

In Section~(\ref{sec-d6bosonicgaugetheories2-forms}) we discussed the
possibility of having (anti-)self dual 2-forms and we found that this can be
described by the tensor $\zeta^M{}_N$. We could ask the same question now in
the context of a gauged theory with massive deformations. The (anti-)self
duality can again be written as

\begin{equation}
 \zeta^M{}_N(J^N-\zeta^N{}_PJ^P)=0\,,
\end{equation}

\noindent
where now $J^N$ contains the hierarchy field strengths $H^M$. This condition
must be consistent with the equations of motion. After hitting the condition
with a covariant derivative we find the following consistency conditions:
Eq.~(\ref{eq:zetaconstraint1}) and

\begin{equation}
 \zeta^M{}_N(Z^{iN}-\zeta^N{}_PZ^{iP})=0\,.
\end{equation}

\noindent
The $\zeta$-tensor is not predicted by the tensor hierarchy because it cannot
distinguish between (A)SD or non-(A)SD 2-forms. This concept only exists once
equations of motion are defined.

\begin{table}[h!]
\begin{center}
\begin{tabular}{|c|c|c|c|c|}
\hline
 &&&&\\
Potential & Gauge  & Interpretation & St\"uckelberg & Existence \\
& transformation & (field strength dual to) & pair with & \\
&&&&\\
\hline
&&&&\\
$B_M$ & massive & $J_M$ & $Z^i{}_MC_i$ & $\forall M\;:\;Z^i{}_M\neq 0$\\
&&&&\\
\hline
&&&&\\
$Z^{iM}B_M$ & massless & $Z^{iM}J_M$ & ungauged $A^i$ & $\forall i\;:\; Z^{iM}\neq 0$\\
&&&&\\
\hline
&&&&\\
$B_M$ & massless & $J_M$ & none & $\forall M\;:\;Z^{iM}=0$\\
&&&&\\
\hline
&&&&\\
$C_i$ & massive & $a_{ij}F^j$ & $\vartheta_i{}^AD_A$ & $\forall i\;:\;\vartheta_i{}^A\neq 0$\\
&&&&\\
\hline
&&&&\\
$Z^i{}_MC_i$ & massless & $Z^i{}_Ma_{ij}F^j$ & $B_M$ & $\forall M\;:\;Z^i{}_M\neq 0$\\
&&&&\\
\hline
&&&&\\
$C_i$ & massless & $a_{ij}F^j$ &  none & $\forall i\;:\;\vartheta_i{}^A=Z^i{}_M=0$\\
&&&&\\
\hline
&&&&\\
$D_A$ & massive & current $j_A$ of & $Y_A{}^\sharp E_\sharp$ & $\forall A\;:\;Y_A{}^\sharp\neq 0$\\
&& symmetry broken by $V$ &&\\
&&&&\\
\hline
&&&&\\
$\vartheta_i{}^AD_A$ & massless & current $j_A$ of & $C_i$ & $\forall i\;:\;\vartheta_i{}^A\neq 0$\\
&& gauged symmetry &&\\
&&&&\\
\hline
&&&&\\
$D_A$ & massless & current $j_A$ of & none & $\forall A\;:\;Y_A{}^\sharp=\vartheta_i{}^A=0$\\
&& global symmetry &&\\
&&&&\\
\hline
 \end{tabular}
\caption{All the $p\ge 2$ forms of the 6-dimensional tensor hierarchy, their St\"uckelberg properties and physical interpretation.}
\label{table:6DTH}
\end{center}
\end{table}

\begin{table}[h!]
\begin{center}
\begin{tabular}{|c|c|c|c|c|}
\hline
 &&&&\\
Potential & Gauge  & Interpretation & St\"uckelberg & Existence \\
& transformation & (field strength dual to) & pair with & \\
&&&&\\
\hline
&&&&\\
$E_\sharp$ & massive & $\partial V/\partial c^{\sharp}$ & $W_\sharp{}^\flat F_\flat$ & $\forall\sharp\;:\;W_\sharp{}^\flat\neq 0$\\
&&&&\\
\hline
&&&&\\
$Y_A{}^\sharp E_\sharp$ & massless & $Y_{A}{}^{\sharp}\partial V/\partial c^{\sharp}$ & $D_A$ & $\forall A\;:\;Y_{A}{}^{\sharp}\neq 0$\\
&&&&\\
\hline
&&&&\\
$E_\sharp$ & massless & $\partial V/\partial c^{\sharp}$ & none & $\forall\sharp\;:\;W_\sharp{}^\flat=Y_A{}^\sharp=0$\\
&&&&\\
\hline
&&&&\\
$W_\sharp{}^\flat F_\flat$ & massless & enforces constraints & $E_\sharp$ & $\forall\sharp\;:\;W_\sharp{}^\flat\neq 0$\\
&&&&\\
\hline
 \end{tabular}
 \caption*{Table 2: (continued)}
\end{center}
\end{table}

The gauge transformations that leave the action
invariant can be written as

\begin{eqnarray}
 \delta A^i & = & -\mathfrak{D}\Lambda^i-Z^{iM}\Lambda_M\,,\\
&&\nonumber\\
\delta B_M & = & \mathfrak{D}\Lambda_M+2d_{M\,ij}\left(\Lambda^iF^j+\frac{1}{2}A^i\wedge\delta A^j\right)-Z_M{}^i\Lambda_i+\Delta B_M\,,\\
&&\nonumber\\
\delta C_i & = & \mathfrak{D}\Lambda_i+2d_{N\,ij}\Lambda^jJ^N-2d_{N\,ij}\Lambda^N\wedge F^j\nonumber\\
&&\nonumber\\
&&-2d_{N\,ij}B^N\wedge\delta A^j-\tfrac{2}{3}d^N{}_{ij}d_{N\,kl}A^{jk}\wedge\delta A^l\,.\label{eq:threeformgauge}
\end{eqnarray}

\noindent
To prove this we only need the following Noether identities associated to the invariance under gauge transformations whose parameters are, respectively $\Lambda^i$, $\Lambda^M$ and $\Lambda_i$,

\begin{eqnarray}
  \mathfrak{D}\star\frac{\widetilde{\delta S}}{\delta A^{i}}+\vartheta_i{}^Ak_A{}^x\star\frac{\delta S}{\delta\phi^x}
+2d^M{}_{ij}F^j\wedge\star\frac{\delta S}{\delta B^M}+2d_{M\,ij}J^M\wedge\frac{\delta S}{\delta C_j}=0\,,\\
\nonumber\\
 \mathfrak{D}\star\frac{\delta S}{\delta B^M}-Z^i{}_M\star\frac{\widetilde{\delta S}}{\delta A^{i}}
-2d_{M\,ij}F^i\wedge\frac{\delta S}{\delta C_j}=0\,,\\
\nonumber\\
 \mathfrak{D}\frac{\delta S}{\delta C_i}-Z^{iM}\star\frac{\delta S}{\delta B^M}=0\,.
\end{eqnarray}

\noindent
We note that these gauge transformations are exactly those of the hierarchy except for the 3-form gauge transformation Eq.~(\ref{eq:threeformgauge}) which can be written as

\begin{equation}
 \delta C_i=\delta_h C_i+2d_{N\,ij}\Lambda^j(J^N-H^N)\,,
\end{equation}
 
\noindent
in which $\delta_hC_i$ (together with the 1-form $\delta A^i$ and 2-form gauge transformations $\delta B_M$) is the gauge transformation under which $H^M$ transforms gauge-covariantly.

We end this section by giving an overview in Table~(\ref{table:6DTH}) of the
6-dimensional tensor hierarchy and its physical interpretation. The way in which Table~(\ref{table:6DTH}) should be read is entirely analogous to the 5-dimensional case discussed at the end of Section~(\ref{sec-d5action}).


\section{Discussion}
\label{sec-discussion}

Without making reference to any particular details of a 5- or 6-dimensional
field theory we have constructed the tensor hierarchies for such theories and
the corresponding gauge-invariant actions. We have found the dualities that
relate these two structures.

Our results, together with those of
Refs.~\cite{Bergshoeff:2009ph,Hartong:2009az} reveal a number of generic
features that must be common to all tensor hierarchies:

\begin{enumerate}

\item The field content of a particular tensor hierarchy provides an exhaustive list of
  all possible potentials that one can introduce into a theory. The generic
  tensor hierarchies that we have constructed provide a minimal
  list. Depending on the existence of additional theory-specific constraints
  (as in the $N=1,d=4$ supergravity case), more potentials may be included.

\item In general, the deformation parameters of any field theory\footnote{In
    this list we are obviously leaving aside deformations such as the
    cosmological constant in non-supersymmetric theories, which are unrelated
    to massive or massless gauge symmetries. These deformation parameters do
    not couple to the hierarchy's $p$-form potentials and, therefore, are
    unaccounted for by it.} are of three different kinds:

  \begin{enumerate}
  \item The embedding tensor $\vartheta$, which determines the gauge group and
    gauge couplings.
  \item The St\"uckelberg tensors $Z$ that will determine the couplings
    between $p$-forms and $(p+1)$-forms and between their respective duals, the
    $(\tilde{p}+1)$-  and $\tilde{p}$-forms (with $\tilde{p}=d-p-2$).
  \item The Chern-Simons tensors $d$ which determine the Chern-Simons terms
    in the field strengths and action.
  \end{enumerate}

\item As explained in the introduction, the tensor hierarchy will contain one
  $(d-1)$-form potential (``de-form'') conjugate to each deformation
  parameter. In a democratic formulation, the de-forms will enforce the
  constancy of the corresponding deformation parameters. There may be
  additional top-forms associated to theory-specific constraints which cannot
  be studied in our generic models. It is unclear if there might be additional
  top-forms whose gauge transformations are unconnected to the hierarchy\footnote{
    What is also still an open question is how to construct the tensor
    hierarchy of a theory without vectors such as the type IIB supergravity
    theory.}.

\item These deformation parameters will be subject to four generic kinds of
  constraints:

  \begin{enumerate}
  \item Constraints that enforce the gauge-invariance of all deformation
    tensors: $\delta \vartheta=0\, ,\,\delta Z=0\, ,\,\delta d =0$. The first
    of these is the standard \textit{quadratic constraint} of the literature.

  \item Orthogonality constraints between the embedding tensor and the first
    St\"uckelberg tensor $\vartheta \cdot Z=0$ and between each St\"uckelberg
    tensor and the next one $Z\cdot Z^{\prime}=0$.

  \item Constraints that relate the $X$ matrices with the Chern-Simons and
    St\"uckelberg or embedding tensors: $X\sim Z\cdot d=0$. The so-called
    linear or representation constraint of the 4-dimensional theories can be
    viewed as an example of this kind of constraints.

  \item Constraints between products of Chern-Simons tensors $d\cdot d=0$. 

  \end{enumerate}

\item As explained in the introduction, the tensor hierarchy will contain a
  top-form potential conjugate to each of the constraints satisfied by the
  deformation tensors. In a democratic formulation, these top-form potentials
  will enforce the corresponding constraints.

\item In $d$-dimensions, a gauge-invariant action for the physical theory can
  be constructed using just the forms of rank 1 to $[d/2]$ (i.e.~2 in $d=4,5$ and
  3 in $d=6,7$ etc.). The gauge transformations will be identical to those of
  the tensor hierarchy up to duality relations. These duality relations are
  essential to relate the tensor hierarchy to the physical theory and fix the
  way all the fields appear in the Lagrangian except for those scalars that
  are not participating in isometry currents.

\end{enumerate}

A tensor hierarchy together with a set of duality relations for its field
strengths (a structure called duality hierarchy in
Ref.~\cite{Bergshoeff:2009ph}) is clearly a powerful tool to construct the
most general bosonic field theory in a particular dimension. This can then be
used as a starting point for the construction of more general supergravity
theories by subsequently supersymmetrizing the hierarchy.


\section*{Acknowledgments}

This work was supported in part by the Swiss National Science Foundation and
the ``Innovations- und Kooperationsprojekt C-13'' of the Schweizerische
Universit\"atskonferenz SUK/CUS. JH wishes to thank the Instituto de
F\'{i}sica Te\'{o}rica of the Universidad Aut\'{o}noma de Madrid for its
hospitality. This work has been supported in part by the Spanish Ministry of
Science and Education grant FPA2006-00783, the Comunidad de Madrid grant
HEPHACOS P-ESP-00346 and by the Spanish Consolider-Ingenio 2010 program CPAN
CSD2007-00042.  Further, TO wishes to express his gratitude to
M.M.~Fern\'andez for her permanent support.

\appendix

\section{Conventions and some formulae}
\label{app-conventions}

We use mostly-minus signature both in 5- and 6-dimensions.

$p$-forms are normalized as follows

\begin{equation}
\omega \equiv \tfrac{1}{p!}\omega_{\mu_{1}\cdots \mu_{p}}
dx^{\mu_{1}}\wedge\cdots \wedge dx^{\mu_{p}}\, .  
\end{equation}

The exterior product of a $p$-form $\omega$ and a $q$-form $\eta$ is 

\begin{equation}
\omega\wedge \eta \equiv \tfrac{1}{p!q!}
\omega_{\mu_{1}\cdots \mu_{p}}\eta_{\nu_{1}\cdots \nu_{q}} dx^{\mu_{1}}\wedge
\cdots \wedge dx^{\mu_{p}}\wedge dx^{\nu_{1}}\wedge \cdots \wedge dx^{\nu_{q}}\, ,  
\end{equation}

\noindent 
so, its components are

\begin{equation}
(\omega\wedge \eta)_{\mu_{1}\cdots \mu_{p+q}} = 
\frac{(p+q)!}{p!q!} 
\omega_{[\mu_{1}\cdots \mu_{p}}\eta_{\mu_{p+1}\cdots \mu_{p+q}]}\, .
\end{equation}

The exterior derivative of a $p$-form $\omega$ is 

\begin{equation}
d\omega \equiv \tfrac{1}{p!}\partial_{\nu}\omega_{\mu_{1}\cdots \mu_{p}}
dx^{\nu}\wedge dx^{\mu_{1}}\wedge\cdots \wedge dx^{\mu_{p}}\, ,  
\end{equation}

\noindent
so, its components are

\begin{equation}
(d\omega)_{\mu_{1}\cdots \mu_{p+1}} = 
(p+1)\partial_{[\mu_{1}}\omega_{\mu_{2}\cdots \mu_{p+1}]}\, .  
\end{equation}

The $d$-dimensional volume form is, with mostly minus signature, 

\begin{equation}
\sqrt{|g|}d^{d}x \equiv \frac{(-1)^{d-1}}{d!\sqrt{|g|}}
\epsilon_{\mu_{1}\cdots \mu_{d}}dx^{\mu_{1}}\wedge \cdots dx^{\mu_{d}}\, ,  
\end{equation}

\noindent
where we have defined the completely antisymmetric symbol such that (in curved
indices)

\begin{equation}
\epsilon^{01\cdots (d-1)}=+1\, ,
\hspace{1cm}
\epsilon_{01\cdots (d-1)}= g \equiv \mathrm{det} g = (-1)^{d-1}|g|\, .  
\end{equation}

The components of the Hodge dual of a $p$-form $\omega$ are defined by 

\begin{equation}
(\star \omega)_{\mu_{1}\cdots \mu_{d-p}}  \equiv 
\frac{1}{p!\sqrt{|g|}}
\epsilon_{\mu_{1}\cdots \mu_{d-p}\nu_{1}\cdots \nu_{p}}\omega^{\nu_{1}\cdots
  \nu_{p}}\, ,
\end{equation}

\noindent
so 

\begin{equation}
\star \omega =   
\frac{1}{p!(d-p)!\sqrt{|g|}}
\epsilon_{\mu_{1}\cdots \mu_{d-p}\nu_{1}\cdots \nu_{p}}
\omega^{\nu_{1}\cdots \nu_{p}} dx^{\mu_{1}}\wedge\cdots \wedge dx^{\mu_{d-p}}\, .  
\end{equation}

\noindent
Then, for $p$-forms $\omega$ in $d$ dimensions, with mostly minus signature,

\begin{equation}
  \star^{2}\omega = (-1)^{d-1+p(d-p)}\,  \omega\, .  
\end{equation}

\noindent
It follows that for 3-forms $H$ in 6 dimensions we have $\star^{2}=+1$ so that we can have real self-
and anti-self-dual 3-forms $H^{\pm}$

\begin{equation}
H^{\pm}\equiv \tfrac{1}{2}(H\pm\star H)\, ,
\hspace{1cm}
\star H^{\pm}=\pm H^{\pm}\, .  
\end{equation}

A $d$-form $\Omega$ in $d$-dimensions is always
proportional to the volume form. We can always write

\begin{equation}
  \begin{array}{rcl}
\Omega 
& = &  
K \sqrt{|g|}\, d^{d}x\, ,\\
& & \\
K 
& = & 
{\displaystyle\frac{1}{d!\sqrt{|g|}}} \epsilon^{\mu_{1}\cdots \mu_{d}}
\Omega_{\mu_{1}\cdots \mu_{d}}\, .\\
\end{array}
\end{equation}

\noindent
Using this property, we find the following formulae in $d$ dimensions

\begin{eqnarray}
\star R 
& = & 
(-1)^{d-1}R \sqrt{|g|}d^{d}x\, ,\\
& & \nonumber \\
d\phi \wedge \star d\phi 
& = & 
(\partial\phi)^{2} \sqrt{|g|}\, d^{d}x\, ,\\
& & \nonumber \\
F\wedge \star F  
& = & 
\tfrac{(-1)^{d-1}}{2} F^{2} \sqrt{|g|}\, d^{d}x\, ,\\
& & \nonumber \\
H\wedge \star H
& = & 
\tfrac{1}{3!} H^{2} \sqrt{|g|}\, d^{d}x\, ,\\
& & \nonumber \\
H\wedge \star \tilde{H}
& = & 
\tfrac{1}{3!} H_{\mu\nu\rho}
(\star \tilde{H})^{\mu\nu\rho} \sqrt{|g|}\, d^{d}x\, .
\end{eqnarray}


\section{Summary of the general 5-dimensional tensor hierarchy }
\label{app-5d}


\subsection{Deformation tensors and  constraints}

The deformation tensors of 5-dimensional field theories are $\vartheta_{I}{}^{A}$, 
$Z^{IJ}=Z^{[IJ]}$ and $C_{IJK}=C_{(IJK)}$. They
are subject to the constraints

\begin{eqnarray}
Q_{IJ}{}^{A} 
& = & 
=
-\vartheta_{I}{}^{B} Y_{B\, J}{}^{A}
= 
-\vartheta_{I}{}^{B}
(\vartheta_{J}{}^{C}f_{BC}{}^{A}
-T_{B\, J}{}^{K}\vartheta_{K}{}^{A})\, , 
\\
& & \nonumber \\
Q_{I}{}^{JK}
& = &  
-\vartheta_{I}{}^{A}
Y_{A}{}^{JK} 
=
2\vartheta_{I}{}^{A}
T_{A\, L}{}^{[J}Z^{K]L}\, ,
\\
& & \nonumber \\
Q_{IJKL}
& = & 
-\vartheta_{I}{}^{A}Y_{A\, JKL}
=
3\vartheta_{I}{}^{A}
T_{A\, (J}{}^{M}C_{KL)M}\, ,  
\end{eqnarray}

\noindent
which express the gauge-invariance of the deformation tensors and 

\begin{eqnarray}
Q^{AI} 
& = &  
\vartheta_{J}{}^{A}Z^{JI}\, ,
\\
& & \nonumber \\
Q_{JK}{}^{I} 
& = &  
X_{(JK)}{}^{I}- Z^{IL} C_{JKL}\, .
\end{eqnarray}


\subsection{Field strengths and Bianchi identities}

The tensor hierarchies of general 5-dimensional bosonic field theories have
1-forms $A^{I}$, 2-forms $B_{I}$, 3-forms $C_{A}$, 4-forms
$D^I{}_{B}$, $D_{IJ}$, $D^{IJK}$ and 5-forms $E^{IJ}{}_{A}$, $E^{I}{}_{JK}$, 
$E^{I\,JKL}$, $E_{A\, I}$ and $E^{IJ}{}_{K}$. The field strengths of the 1-, 2-, 3- and
4-form fields are given by

\begin{eqnarray}
F^{I} 
& = & 
dA^{I}
+\tfrac{1}{2}X_{JK}{}^{I}A^{JK}  +Z^{IJ}B_{J}\, ,
\\
& & \nonumber \\
H_{I} 
& = & 
\mathfrak{D}B_{I}+C_{IJK}A^{J}\wedge dA^{K}
+\tfrac{1}{3} C_{IM[J}X_{KL]}{}^{M}A^{JKL}  
+\vartheta_{I}{}^{A}C_{A}\, ,
\\
& & \nonumber \\
G_{A} 
& = & 
\mathfrak{D}C_{A} 
+T_{A\, K}{}^{I} 
\left[ 
  (F^{K}-\tfrac{1}{2}Z^{KL}B_{L})\wedge B_{I}
+\tfrac{1}{3}C_{ILM}A^{KL}\wedge dA^{M} 
\right.
\nonumber \\
& & \nonumber \\
& & 
\left.
+\tfrac{1}{12}C_{ILP}X_{MN}{}^{P}A^{KLMN} 
\right] 
+Y_{A}{}^{IJ}D_{IJ}+Y_{A\, I}{}^{B}D_{B}{}^{I}+Y_{A\, IJK}D^{IJK}\, ,
\\
& & \nonumber \\
K^I{}_{B}
& = & 
\mathfrak{D} D^I{}_{B}
+(F^{I}-Z^{IL}B_{L})\wedge C_{B}
+\tfrac{1}{12}T_{B\, J}{}^{M}C_{KML}A^{IJK} \wedge dA^{L}
\nonumber \\
& & \nonumber \\
& & 
+\tfrac{1}{60}T_{B\, J}{}^{N}C_{KPN}X_{LM}{}^{P}A^{IJKLM}
+W_{B}{}^{I}{}_{KJ}{}^{D} E^{KJ}{}_{D}
-Z^{IJ}E_{B\, J}
-T_{B\, K}{}^{J}E_{J}{}^{IK}
\nonumber \\
& & \nonumber \\
& & 
-Y_{B}{}^{JK}E^{I}{}_{JK}\, ,
\\
& & \nonumber \\
K_{IJ}
& = & 
\mathfrak{D} D_{IJ}
-\left[H_{[I}-\tfrac{1}{2}\mathfrak{D}B_{[I}\right]\wedge B_{J]}
+2X_{K[I}{}^{L}E^{K}{}_{J]L}
-C_{KL[I}E^{KL}{}_{J]}
\nonumber \\
& & \nonumber \\
& & 
-\vartheta_{[I|}{}^{A}E_{A\, |J]}\, ,
\\
& & \nonumber \\
K^{IJK}
& = & 
\mathfrak{D} D^{IJK}
+\tfrac{1}{3}A^{(I}\wedge dA^{JK)}
+\tfrac{1}{4}X_{LM}{}^{(K} A^{I|LM}\wedge dA^{|J)}
\nonumber \\
& & \nonumber \\
& & 
+\tfrac{1}{20}X_{LM}{}^{(J}X_{NP}{}^{K}
A^{I)LMNP}
+3 X_{LM}{}^{(I|}E^{L\, |JK)M} 
+Z^{L(I}E_{L}{}^{JK)}\, ,
\end{eqnarray}

\noindent
and are related by the Bianchi identities

\begin{eqnarray}
\mathfrak{D}F^{I} 
& = & 
Z^{IJ} H_{J}\, ,
\\
& & \nonumber \\
\mathfrak{D}H_{I} 
& = & 
C_{IJK}F^{JK}+\vartheta_{I}{}^{A}G_{A}\, ,
\\
& & \nonumber \\
\mathfrak{D}G_{A}
& = &   
T_{A\, K}{}^{I}F^{K}\wedge H_{I}
+Y_{A}{}^{IJ}K_{IJ}
+Y_{A\, I}{}^{B}K^I{}_{B}
+Y_{A\, IJK}K^{IJK}\, . 
\end{eqnarray}


\subsection{Duality relations}

\begin{eqnarray}
H_{I} & = & a_{IJ}\star F^{J} \, ,
\\
& & \nonumber \\
G_{A}
& = & 
\star j_{A}\, ,
\\
& & \nonumber \\
K_{\sharp} 
& = & 
\star \frac{\partial V}{\partial c^{\sharp}}\, .
\end{eqnarray}


\section{Summary of the general 6-dimensional tensor hierarchy }
\label{app-6d}


\subsection{Deformation tensors and  constraints}

The deformation tensors of 6-dimensional field theories are $\vartheta_{i}{}^{A}$, $Z^{iM}$ and $d_{M\, ij}= d_{M\, (ij)}$. They
are subject to the constraints

\begin{eqnarray}
Q_{ji}{}^{A}
& \equiv & 
-\vartheta_{j}{}^{B}Y_{B\, i}{}^{A} 
= 
-\vartheta_{j}{}^{B}
(f_{BC}{}^{A}\vartheta_{i}{}^{C}-T_{B\, i}{}^{k}\vartheta_{k}{}^{A})\, ,
\\
& & \nonumber \\
Q_{j}{}^{iM}
& \equiv & 
-\vartheta_{j}{}^{A} Y_{A}{}^{iM}
=
-\vartheta_{j}{}^{A}
(T_{A\, k}{}^{i} Z^{kM} +T_{A\, N}{}^{M}Z^{iN})\, ,
\\
& & \nonumber \\
Q_{k\, M\, ij}
& \equiv & 
-\vartheta_{k}{}^{A}
Y_{A\, M\, ij}
=
\vartheta_{k}{}^{A}
(2T_{A\, (i|}{}^{l}d_{M\, |j)l} +T_{A\, M}{}^{N}d_{N\, ij})\, ,
\end{eqnarray}

\noindent
associated to their gauge-invariance and, furthermore, to the constraints

\begin{eqnarray}
Q^{AM} 
& \equiv  & 
\vartheta_{i}{}^{A}Z^{iM}\, ,
\\
& & \nonumber \\
Q^{ij} 
& \equiv & 
Z^{iM}Z^{j}{}_{M} \, ,
\\
& & \nonumber \\
Q_{jk}{}^{i}
& \equiv  & 
X_{(jk)}{}^{i} -Z^{iM}d_{M\, jk}\, ,
\\
& & \nonumber \\
Q_{i\, MN}
& \equiv & 
X_{i\, MN} - 4Z^j{}_{[M}d_{N] ij}\, ,
\\
& & \nonumber \\
Q_{ijk\, l} 
& \equiv  & 
d_{M\, (ij}d^{M}{}_{k)l}\, .
\end{eqnarray}


\subsection{Field strengths and Bianchi identities}

The tensor hierarchies of general 6-dimensional bosonic field theories have
1-forms $A^{i}$, 2-forms $B_{M}$, 3-forms $C_{i}$, 4-forms $D_{A}$, three types
of 5-forms $E^{i}{}_{A},E_{iM},E^{M\, ij}$ and eight types of 6-forms (that we will only refer to collectively as $F_\flat$). The field
strengths of the 1- to 5-form potentials are given by

\begin{eqnarray}
F^{i} 
& = & 
dA^{i}+\tfrac{1}{2}X_{jk}{}^{i}A^{jk}  +Z^{iM}B_{M}\, ,\\
& & \nonumber \\  
H_{M} 
& \equiv & 
\mathfrak{D}B_{M}+ d_{M\, jk}[A^{j}\wedge dA^{k}
+\tfrac{1}{3}X_{lm}{}^{k}A^{jlm}]  
-Z^i{}_MC_{i}\, ,
\\
& & \nonumber \\
G_{i}
& = & 
\mathfrak{D}C_{i}
+2d^{N}{}_{ip}
\left[
(F^{p} -\tfrac{1}{2}Z^{pM}B_{M})\wedge B_{N}
+\tfrac{1}{3}d_{N\, jk} A^{pj}\wedge dA^{k}
+\tfrac{1}{12} X_{jk}{}^{n}d_{N\, ln}A^{pjkl}
\right]
\nonumber \\
& & \nonumber \\
& & 
+\vartheta_{i}{}^{A} D_{A}\, ,
\\
& & \nonumber \\
K_{A} 
& = & 
\mathfrak{D}D_{A} 
+T_{A}{}^{MN}(H_{M}-\tfrac{1}{2}\mathfrak{D}B_{M})\wedge B_{N}
\nonumber \\
& & \nonumber \\
& & 
+T_{A\, k}{}^{p}
\left[
(F^{k}-Z^{kM}B_{M})\wedge C_{p}
-\tfrac{1}{6}
d^{M}{}_{jp}d_{M\, lm} A^{jkl}\wedge dA^{m}
+\tfrac{1}{30}X_{lm}{}^{q}d^{M}{}_{jq}d_{M\, pn} A^{jklmn}
\right]
\nonumber \\
& & \nonumber \\
& & 
+Y_{A\, i}{}^{B} E^{i}{}_{B}
+Y_{A}{}^{iM}E_{iM}
+Y_{A\, M\, ij}E^{M\, ij}\, ,
\\
& & \nonumber \\
L^{i}{}_{B}
& = & 
\mathfrak{D}E^{i}{}_{B}-F^{i}\wedge D_{B}
+\tfrac{1}{30} T_{B\, k}{}^{n}d^{N}{}_{jm}d_{N\, ln}A^{ijkl}\wedge dA^{m} 
+\tfrac{1}{80}  T_{B\, k}{}^{p} X_{lm}{}^{q} 
d^{N}{}_{jq}d_{N\, pn}A^{ijklmn}
\nonumber \\
& & \nonumber \\
& & 
+\frac{\partial Q^\flat}{\partial\vartheta_i{}^B}F_\flat\, ,\label{eq:LiB}
\\
& & \nonumber \\
L_{iM}
& = &
\mathfrak{D}E_{iM}
+(H_{M}-\mathfrak{D}B_{M})\wedge C_{i} 
-B_{M}\wedge (G_{i}-\vartheta_{i}{}^{B}D_{B}) 
-\tfrac{1}{2}Z^{j}{}_{M}C_{ij}
\nonumber \\
& & \nonumber \\
& & 
+d^N{}_{ij}F^{j}\wedge B_{MN}
+\tfrac{1}{3}d^N{}_{ij}Z^{jP}B_{MNP}
+\frac{\partial Q^\flat}{\partial Z^{iM}}F_\flat\, ,\label{eq:LiM}
\\
& & \nonumber \\
L_{M}{}^{ij}
& = & 
\mathfrak{D}E_{M}{}^{ij}
-F^{ij}\wedge B_{M} +Z^{iN}F^{j} \wedge B_{MN}
-\tfrac{1}{3}Z^{iN}Z^{jP}B_{MNP}
\nonumber \\
& & \nonumber \\
& & 
-\tfrac{1}{2} d_{M\, kl}A^{ik}\wedge dA^{jl} 
-\tfrac{2}{15} X_{kl}{}^{n}d_{M\, nm} A^{iklm}\wedge dA^{j}
-\tfrac{1}{5}X_{kl}{}^{j} d_{M\, nm} A^{ikln}\wedge dA^{m}
\nonumber \\
& & \nonumber \\
& & 
-\tfrac{1}{18}X_{kl}{}^{j} X_{np}{}^{q} d_{M\, mq}A^{iklmnp}
+\frac{\partial Q^\flat}{\partial d^M{}_{ij}}F_\flat\,.\label{eq:LMij}
\end{eqnarray}

\noindent
These field strengths are related by the following Bianchi identities

\begin{eqnarray}
\mathfrak{D}F^{i} 
& = & 
Z^{iM} H_{M}\, ,
\\
& & \nonumber \\
\mathfrak{D} H_{M} 
& = & 
d_{M\, ij} F^{ij} 
-Z^i{}_{M} G_{i}\, ,
\\
& & \nonumber \\
\mathfrak{D}G_{i} 
& = & 
2 d^{M}{}_{ij}F^{j}\wedge H_{M}  +\vartheta_{i}{}^{A}K_{A}\, ,
\\
& & \nonumber \\
\mathfrak{D}K_{A} 
& = &  
T_{A\, j}{}^{k} F^{j}\wedge G_{k}   
-\tfrac{1}{2} T_{A}{}^{MN}H_{MN}
\nonumber \\
& & \nonumber \\
& & 
+Y_{A\, i}{}^{B}L^{i}{}_{B}
+Y_{A}{}^{iM} L_{iM}
+Y_{A}{}^{M}{}_{ij}L_{M}{}^{ij}\, .\label{eq:BianchiKA}
\end{eqnarray}


\subsection{Duality relations}

\begin{eqnarray}
H_{\Lambda} 
& = & 
J_{\Lambda} 
=  
b_{\Lambda\Sigma}\star H^{\Sigma} +c_{\Lambda\Sigma}H^{\Sigma}\, ,
\\
& & \nonumber \\  
G_{i} & =  & a_{ij}\star F^{j}\, ,
\\
& & \nonumber \\  
K_{A}
& = & 
\star j_{A}\, ,
\\
& & \nonumber \\
L_{\sharp}
& = & 
 \star \frac{\partial V}{\partial c^{\sharp}}\, .
\end{eqnarray}


\end{document}